\documentclass[11pt,tightenlines,aps,prd,groupedaddress,nofootinbib,nobibnotes,longbibliography,notitlepage]{revtex4-1}
\pdfoutput=1

\usepackage{graphicx}
\usepackage{amsfonts}
\usepackage{amssymb}
\usepackage{amsbsy}
\usepackage{amsmath}
\usepackage{mathtools}
\usepackage{latexsym}
\usepackage{bm}
\usepackage{color}
\usepackage{hyperref}
\usepackage[margin=2cm]{geometry}
\usepackage{comment}
\usepackage[makeroom]{cancel}
\usepackage[dvipsnames]{xcolor}

\usepackage{stackengine}
\stackMath
\newcommand\tenq[2][1]{%
 \def\useanchorwidth{T}%
  \ifnum#1>1%
    \stackon[0pt]{\tenq[\numexpr#1-1\relax]{#2}}{\scriptscriptstyle\sim}%
  \else%
    \stackon[1pt]{#2}{\scriptscriptstyle\sim}%
  \fi%
}

\def\r{\right}

\newcommand{\de}{\mbox{d}}
\newcommand{\lf}{\left}
\newcommand{\rg}{\right}
\newcommand{\be}{\begin{equation}}
\newcommand{\ee}{\end{equation}}

\newcommand{\bea}{\begin{eqnarray}}
\newcommand{\eea}{\end{eqnarray}}

\numberwithin{equation}{section}

\begin{document}

\title{\Large Arrows of time in bouncing cosmologies}

\homepage[Chapter contributed to the volume ``Time and timelessness in fundamental physics and cosmology'', edited by S.~De~Bianchi, M.~Forgione, and L.~Marongiu, {\it Fundamental Theories of Physics} series (Springer)]{}

\author{Marco de Cesare}
\email{marco.decesare@na.infn.it}

\affiliation{Scuola Superiore Meridionale, Largo San Marcellino 10, 80138 Napoli, Italy}
\affiliation{INFN, Sezione di Napoli, Italy}


\begin{abstract}
Different approaches to quantum gravity, such as loop quantum cosmology and group field theory, predict the resolution of the initial cosmological singularity via a `bounce': a regular spacetime region that connects the expanding branch of the universe to a contracting branch. The cosmological arrow of time, which by definition points in the direction of cosmic expansion, is reversed at the bounce. Nonetheless, it is still possible to discriminate between the two branches by considering different arrows, as defined for instance by the growth of perturbations.
After reviewing general aspects of the time arrow problem in cosmology, we examine the properties of different arrows of time in bouncing cosmologies, focusing on the loop quantum cosmology bounce as a case study.
These issues are examined in detail for an exact solution to the effective Friedmann equations of loop quantum cosmology with pressureless dust and a cosmological constant, which is a simplified version of the $\Lambda$CDM bounce scenario.
\end{abstract}

\maketitle



\section{Arrows of time and their cosmological origin}\label{Sec:1}
The physical world offers a wealth of examples of phenomena taking place in a definite temporal direction, which are either very rarely or never spontaneously reversed. Despite the reversibility
of fundamental physical laws at a microscopic level,\footnote{More precisely, local relativistic quantum field theories are invariant under CPT, i.e.~the combination of time reversal (T), charge conjugation (C), and parity (P). This ensures the absence of a microscopic arrow of time.}
the evolution of macroscopic physical systems is typically characterised by an unequivocal directionality of time that allows us to discriminate the past from the future.
 This basic fact permeates virtually every aspect of our experience. For instance, we never observe eggs uncracking, footsteps appear in the sand after the retreating tide, or black holes disappear into uncollapsed matter. Several distinct {\it arrows of time} can be identified in nature, based on intrinsic temporal asymmetries observed in different classes of physical phenomena \cite{Hawking:1985af,Gell-Mann:1991kdm}:
\begin{itemize}
\item The thermodynamical arrow of time, pointing in the direction of increasing entropy;
\item The psychological arrow of time, according to which we can only remember our past but not our future;
\item The electrodynamical arrow of time, explaining why only retarded solutions of the Maxwell equations are observed but never advanced ones;
\item The cosmological arrow of time, pointing in the direction of cosmic expansion. In standard cosmology based on general relativity, the arrow always points away from the initial Big Bang singularity if the universe is flat or open, whereas in a closed universe the arrow may reverse its direction at late times if the cosmological constant is smaller than a certain critical value. The situation is different in bouncing cosmologies, where the cosmological arrow is reversed at the bounce and points away from it on either side;
\item The arrow of time specified by the direction in which density perturbations grow, which ultimately determines the formation of structures in the universe. 
\end{itemize}

Further arrows of time are also defined by, e.g., state vector reduction in the measurement of a quantum system, diffusive and radiative processes in astrophysics, biological processes, and so on \cite{ELLIS2013242}. Although conceptually distinct, the different arrows of time are not independent of one another. For instance, the psychological arrow clearly depends on the thermodynamical one, as biological systems obey the laws of thermodynamics. The electrodynamical arrow is determined by a boundary condition of no incoming radiation in the past, which in the Wheeler-Feynman absorber theory is traced back to the thermodynamic properties of the absorbing medium \cite{Hawking:1985af,Zeh_2011,Forgione:2022nms}.
More in general, if the physical descriptions of systems are organized hierarchically, ranging from the smallest to the largest scales through successive levels of emergence, it is possible to identify relations between different levels that proceed in both a bottom-up and a top-down manner \cite{ELLIS2013242}. 

We stress that, even for physical systems which are governed by time-symmetric equations, in general we do not expect that solutions enjoy the same symmetry as the equations. Rather, solutions would come in pairs that are the image of one another under time reversal. The reason why only certain such solutions are physically realized, but not their time-reversed counterpart, ultimately hinges on the boundary conditions imposed on the system at hand, which are typically determined by the environment~\cite{Zeh_2011}.

The arrows of time corresponding to different levels of description of the physical reality are locally determined, and thus may in principle point in opposite directions in different spacetime regions.\footnote{This possibility, though speculative, is logically consistent and has been explored in Ref.~\cite{Schulman:1999kg} in a toy model with two weakly-coupled systems with opposite-pointing thermodynamical time arrows, realized by means of two-time boundary conditions (see also Ref.~\cite{Zeh_2005} for a critical discussion).} However, each of the arrows happen to consistently point in the same direction across the observable universe: this constitutes the so-called {\it locality issue} \cite{ELLIS2013242}.  The remarkable agreement between the different arrows of time as well as their spatiotemporal coherence over vastly different scales suggest that they may have a common origin.
Thus, one may speculate that the cosmological arrow of time in the early universe played the role of a {\it global master arrow}, from which all other arrows can be ultimately derived~\cite{ELLIS2013242,Kiefer_2011}.

According to an early speculative proposal by Gold \cite{Gold_1962}, the thermodynamical arrow of time may coincide with the cosmological one {\it at all times}. In a closed universe, this would imply the reversal of the thermodynamical arrow at the turnaround point, where the universe enters the recollapse phase. All physical processes would then run in the opposite time direction, and all memories and records created during the expanding phase would be erased. A realization of this scenario necessarily requires low entropy boundary conditions both at the Big Bang and Big Crunch. However, this proposal has not been generally accepted and faced severe criticism~\cite{Penrose:1979azm,Zeldovich:1983cr}. 
It was argued by Hawking~\cite{Hawking:1985af} that a scenario similar to the Gold universe could be naturally realized in the no-boundary proposal for the path integral in quantum cosmology, although Ref.~\cite{Page:1985ei} later showed that this is not necessarily the case. In the Wheeler-De Witt approach to quantum cosmology, a natural choice of `time' variable is represented by (the logarithm of) the cosmic scale factor; hence, it is natural to impose boundary conditions in the limit where the universe is small (i.e, the `far past' for such a choice of time variable), which corresponds to both initial and final cosmological singularities in a closed universe. Imposing low-entropy boundary conditions in the limit where the scale factor is small then leads to a thermodynamical time arrow that agrees with the cosmological one and reverses at the point of recollapse \cite{Kiefer:1994gp}.

It has been pointed out by Penrose \cite{Penrose:1979azm,Penrose:1988mg,Penrose:2006zz} that the entropy of the observable universe is much smaller by several orders of magnitude compared to its maximal value. The latter can be estimated as the Bekenstein-Hawking entropy of a spherical black hole with the same total mass as the observable universe. In a universe with baryon number $10^{80}$, the maximal entropy is about $S_{\rm max}\approx10^{123}$ in natural units\footnote{This is actually a lower bound for the maximal entropy since dark matter has not been taken into account \cite{Penrose:2006zz}.}, whereas the entropy of the cosmic microwave background (CMB) radiation is $S\approx10^{88}$. The extremely small value of $S$ relative to $S_{\rm max}$ implies that the universe must have started in a rather special state at the Big Bang. Moreover, CMB observations show that, while matter was in a thermal (i.e., maximal entropy) state at the time of recombination, the gravitational field was extremely uniform and thus in a low entropy state.\footnote{Although a general definition of gravitational entropy is still not available (with the exception of specific systems, such as black holes and homogeneous cosmological models \cite{Bardeen:1973gs,Davies:1988dk,Jacobson:2003wv}), a qualitative relation between entropy increase and gravitational clumping has been pointed out by Penrose \cite{Penrose:1979azm,Penrose:1988mg}. Qualitatively, the thermodynamic behaviour of a system of particles that interact gravitationally is opposite to that of a gas, due to the long range nature of the gravitational force. Spatially uniform distributions of matter correspond to low entropy states for the gravitational field. As matter particles start to clump together to form gravitationally bound structures, the entropy grows. Eventually, black holes would form and the entropy reaches its maximum value when the totality of matter collapses into a single black hole. Since the absence of clumping roughly corresponds to the vanishing of the Weyl tensor, Penrose conjectured that the latter may be a measure of entropy \cite{Penrose:1979azm} (see also \cite{Husain:1988jf}). However, this proposal faced some difficulties \cite{BONNOR1987305,Goode:1992pp} (see also \cite{Clifton:2013dha}).}
This motivated the conjecture ({\it Weyl curvature hypothesis}) that in our universe the Weyl curvature vanishes in the approach to initial spacetime singularities, and therefore gravitational degrees of freedom are not excited at the Big Bang \cite{Penrose:1979azm,Penrose:1988mg,Penrose:2006zz}. On the other hand, final spacetime singularities are unconstrained. This hypothesis ensures the smoothness of the initial state of the universe, thus providing a low-entropy past boundary condition that explains the existence of a thermodynamical time arrow pointing in the direction of cosmic expansion \cite{Penrose:1994de}.
We note that, in general, solutions to the Einstein equations would display chaotic Belinskii-Khalatnikov-Lifshitz (BKL) oscillatory behaviour in the approach to the initial cosmological singularity, which implies a highly anisotropic and inhomogeneous initial state \cite{Belinsky:1982pk}. Thus, the Weyl curvature hypothesis constitutes indeed a strong constraint on the initial state.

The Weyl curvature hypothesis represents a particular instance of the so-called {\it Past Hypothesis} \cite{Albert2000-ALBTAC-4}: in order to explain the smallness of the entropy of the universe and its increase with cosmic expansion, it seems necessary to assume a special initial state with an extremely low entropy.  We note that in a universe where the second law of thermodynamics holds, it is impossible to retrodict the initial state. In fact, as a consequence of the time-reversal invariance of the microscopic dynamics, standard arguments from statistical mechanics show that the evolution from the present macrostate will lead to entropy increase {\it both} into the future and into the past \cite{Penrose:1994de,Schiffrin:2012zf}. Since we know from cosmological observations that the entropy of the universe actually decreases towards the past, retrodictions are bound to give incorrect answers. It is therefore necessary to add further theoretical input on the initial state and use the latter to predict the future, rather than attempt to retrodict the past. The status of the Past Hypothesis as a possible solution of the low-entropy puzzle in cosmology has been much debated in the literature; however, some difficulties have been pointed out concerning both its precise formulation and the explanatory power of statistical arguments\footnote{One of the reasons is that the interpretation of the Liouville measure on the space of solutions for cosmological spacetimes as a probability measure faces a number of well-known difficulties \cite{Schiffrin:2012zf,Curiel:2015oea}. A well-defined probability measure would be needed to put on solid ground typicality arguments that are usually made when invoking the Past Hypothesis.} used in the proposal \cite{Price2004-PRIOTO-2,Callender2004-CALTIN-2,Earman2006-EARTPH-3,Gryb:2020thg}.

Alternatively, attempts have been made to explain the low entropy of the observable universe starting from a random initial state as the result of a dynamical process, such as inflation. However, special initial conditions must be assumed at the onset of inflation in order to produce an inflating patch that evolves to a universe like ours; thus, the issue of the exceptionality of the initial state is just shifted to the pre-inflationary epoch \cite{Wald:2005cb,Earman2006-EARTPH-3,Davies:2014}. Yet another approach to explain the initial conditions of the universe is to appeal to quantum cosmology \cite{Vilenkin:1983xq,Hartle:1983ai,Feldbrugge:2017fcc}, where the problem translates into the identification of appropriate boundary conditions in superspace, or, equivalently, of specific integration contours in the path integral over geometries. There are also proposals where entropy grows symmetrically both to the past and to the future of a given low-entropy spatial hypersurface~\cite{Aguirre:2003ck,Carroll:2004pn,Vilenkin:2013rza}.\footnote{An alternative proposal with bidirectional arrows of time representing growth of complexity was made in \cite{Barbour:2014bga}.} Two-times \emph{high-entropy} boundary conditions can also lead to intermediate low entropy states, as shown in Ref.~\cite{Deutsch:2021ylr}.

Going beyond standard cosmology, finding an explanation for the low entropy of the universe remains an open problem also in alternative early universe scenarios. In the remainder of this contribution we will focus specifically on the issue of determining arrows of time in bouncing cosmologies, taking loop quantum cosmology as a case study.

\section{The LQC bounce}\label{Sec:2}

It is well-known that classical general relativity predicts its own breakdown at spacetime singularities \cite{Hawking:1970zqf}. There is a widespread expectation that such singularities must be resolved in a theory of quantum gravity. In the cosmological case, this idea has been realized concretely in the framework of non-perturbative approaches to quantum gravity, such as loop quantum cosmology (LQC) \cite{Ashtekar:2011ni} and group field theory (GFT) \cite{Oriti:2006se,Oriti:2016qtz}, where the initial cosmological ``Big Bang'' singularity is replaced by a ``Big Bounce'' (see also Oriti's contribution in this volume \cite{Oriti:2024elx} and the reviews \cite{Pithis:2019tvp,Gabbanelli:2020lme}).\footnote{See also Ref.~\cite{Barca:2021qdn}.}
In these models, a nonsingular bounce is obtained as a result of quantum gravity effects, which cause the gravitational interaction to turn into a repulsive force at high energy densities, typically of the order of the Planck density. The bounce represents a smooth spacetime region that connects the expanding branch of the universe to a past contracting branch. A similar scenario has been realized also in Pre-Big Bang cosmology in the framework of string theory \cite{Gasperini:2002bn,Gasperini:2023tus,Veneziano:1999ts}, see also Ref.~\cite{Conzinu:2023fth}. Besides quantum gravity, singularity resolution has also been studied in the framework of modified gravity, where early proposals were based on the implementation of the so-called limiting curvature hypothesis \cite{markov1982limiting,Mukhanov:1991zn}, and bouncing cosmologies have been concretely realized more recently in different classes of theories (including, e.g.~scalar-tensor theories, massive gravity, and higher-order modifications of general relativity, see \cite{Easson:2011zy,Cai:2012va,Cai:2012ag,Ijjas:2016tpn,Chamseddine:2016uef,deCesare:2018cts,deCesare:2019suk,Ilyas:2020qja,deCesare:2019pqj,Ilyas:2020qja} and references therein). Bouncing models resolve the classical horizon problem of standard cosmology, and may represent either an alternative to the inflationary mechanism for the generation of primordial cosmological fluctuations  or a complement to inflation that includes the effects of Planck scale physics \cite{Brandenberger:2016vhg,Agullo:2023rqq}.

In this section we focus specifically on the LQC bounce. After reviewing the effective cosmological dynamics, we will consider a concrete model where the energy content consists of pressureless dust and a cosmological constant, and derive an exact solution of the LQC effective Friedmann equations. We will show that no physically meaningful time asymmetry exists at the background level, i.e., both the cosmological and the thermodynamical arrows of time are collinear and point away from the bounce. The dynamics of perturbations and the resulting arrow of time will be studied in the following section.

The effective dynamics for a spatially flat Friedmann-Lema{\^i}tre-Robertson-Walker (FLRW) model are~\cite{Ashtekar:2011ni}
\begin{subequations}
\begin{align}
H^2&=\frac{8\pi G}{3} \rho \lf(1-\frac{\rho}{\rho_c}\rg)~,\label{Eq:FriedEff}\\
\dot{H}&=-4\pi G(\rho+p)\lf(1-2\frac{\rho}{\rho_c} \rg)~.\label{Eq:RayEff}
\end{align}
\end{subequations}
Here $H=\dot{a}/a$ is the Hubble expansion rate, $a$ is the scale factor, $\rho$ and $p$ are the energy density and pressure of matter, respectively. Evolution is defined with respect to proper time $t$ and an overdot denotes time derivative $\dot{\;}=\de/\de t$. The critical energy density is Planckian, $\rho_c\sim\rho_{\rm Pl}$. From Eqs.~\eqref{Eq:FriedEff}, \eqref{Eq:RayEff} follows the continuity equation
\be\label{Eq:ContinuityEquation}
\dot{\rho}+3H (\rho+p)=0~.
\ee 
We assume that $\rho\geq0$ and that the null energy condition is satisfied $\rho+p\geq0$. Equation~\eqref{Eq:ContinuityEquation} then implies that $\rho$ is monotonically decreasing when regarded as a function of $a$. This implies in particular that the maximum of $\rho$ is attained when $a$ is minimized. The bounce takes place in the early universe when the energy density of matter is at a maximum $\rho=\rho_c$~, whereby the Hubble rate vanishes $H=0$ and the scale factor attains its minimum. The initial cosmological singularity is fully resolved.\footnote{Although (harmless) weak curvature singularities may still exist \cite{Singh:2009mz}.} Moreover, Eq.~\eqref{Eq:RayEff} implies that $\dot{H}>0$ in a neighbourhood of the bounce where $\rho_c/2<\rho\leq\rho_c$, whereas $\dot{H}<0$ if $\rho<\rho_c/2$. Standard cosmology is recovered in the regime where $\rho\ll\rho_c$, which is realized both in the late-time and early-time limits.

Note that it is also possible to interpret Eqs.~\eqref{Eq:FriedEff}, \eqref{Eq:RayEff} as the Friedmann equations for an effective fluid with $\tilde{\rho}=\rho(1-\rho/\rho_c)$ and $\tilde{p}=p(1-2\rho/\rho_c)-\rho^2/\rho_c$. The effective fluid violates the null energy condition in a neighbourhood of the bounce, where $\tilde{\rho}+\tilde{p}<0$ and as a consequence $\dot{H}>0$. This allows for the change of sign of the Hubble rate, i.e.~the transition from contraction to expansion. The bounce (and thus the ensuing violation of the null energy condition for the effective fluid) has a very short duration, of the order of the Planck time, which is typical in models motivated from quantum gravity.

Without loss of generality, we can shift the origin of time so that the bounce takes place at $t=0$. With this choice, the universe is in a contracting phase for $t<0$ (where $H<0$), whereas for $t>0$ it is expanding ($H>0$). Also note that at this level the distinction between `past'  and `future' is merely a matter of conventions, and can be reversed by changing the orientation of the time axis. In this and the following section we will investigate whether any {\it physical} time asymmetries emerge, which may enable us to differentiate between the past and the future.

For the sake of concreteness, we consider a model with a cosmological constant $\Lambda>0$ and a pressureless fluid $\rho=\rho_{\rm m}+\Lambda/(8\pi G)$, with $\rho_{\rm m}=\rho_{o}(a_o/a)^3$, where $a_o$ is the scale factor at the bounce. The presence of a positive cosmological constant ensures that the universe is asymptotically de Sitter, and therefore one can define cosmological event horizons (as well as their entropy). The present model is a simplified version of the $\Lambda$CDM bounce scenario in LQC, which includes also the contribution of radiation (which dominates over dust in the early universe) and has been studied earlier in Refs.~\cite{Cai:2014jla,Cai:2016hea}. In turn, the $\Lambda$CDM bounce is a generalization of the LQC matter bounce scenario \cite{Wilson-Ewing:2012lmx} (for a recent review see Ref.~\cite{Agullo:2023rqq}).
Using Eq.~\eqref{Eq:FriedEff} at the bounce, we obtain $\rho_{o}+\Lambda/(8\pi G)=\rho_c$, which can be used to eliminate $\rho_o$. The model is exactly solvable, with scale factor given by\footnote{A generalization of this solution can be obtained in a model with a cosmological constant and a fluid with arbitrary equation of state $w$ (in which case the energy density is $\rho=\rho_{w}+\Lambda/(8\pi G)$, with $\rho_{w}=\rho_{o}(a_o/a)^{3(w+1)}$~). Also in this case there exists an exact solution of the LQC effective Friedmann equation \eqref{Eq:FriedEff} with a very similar structure as \eqref{Eq:ScaleFacSol}, given by $a(t)=a_o \left[1 +\frac{4\pi G \rho_c}{\Lambda} \bigg(\cosh\Big(3(1+w) H_{\infty} t \Big)-1 \bigg) \right]^{\frac{1}{3(1+w)}}$, see Ref.~\cite{Chinaglia:2017wim}.}
\be\label{Eq:ScaleFacSol}
a(t)=a_o \left[1 +\frac{4\pi G \rho_c}{\Lambda} \Big(\cosh\left(3 H_{\infty} t \right)-1 \Big) \right]^{1/3}~,
\ee
where we defined $H_{\infty}\equiv\sqrt{\Lambda \left(1-\Lambda/(8\pi G \rho_c)  \right)/3~ }$~. This solution of the effective Friedmann equations in LQC has been obtained earlier in Ref.~\cite{Chinaglia:2017wim}.

In our universe the cosmological constant is much smaller than the Planck curvature scale $8\pi G \rho_c\sim \ell_{\rm Pl}^{-2}\gg\Lambda$. Therefore, we have approximately $H_{\infty}\approx \sqrt{\Lambda/3}$ for the value of the Hubble rate at late times. From the exact solution \eqref{Eq:ScaleFacSol} we can obtain the Hubble rate as a function of time
\be
H(t)=\frac{H_{\infty}  \sinh\left(3 H_{\infty} t \right)}{(\Lambda/4\pi G\rho_c)+\cosh\left(3 H_{\infty} t \right)-1}~.
\ee
The duration of the bounce, computed as the duration of the time interval where $\dot{H}>0$, is $T\approx \sqrt{2/(3\pi G\rho_c)}\sim t_{\rm Pl}$. Note that in the $\Lambda\to0$ limit, the universe is filled with pressureless dust and thus we recover the LQC matter bounce solution \cite{Wilson-Ewing:2012lmx}
\be
a(t)\approx a_o\lf( 1+6\pi G \rho_c t^2\rg)^{1/3}~,\quad H(t)\approx\frac{t}{(4\pi G \rho_c)^{-1}+\frac{3}{2}t^2}~.
\ee

In all bouncing models, the cosmological arrow of time determined by the background reverses its direction at the bounce, since by definition it always points in the direction of cosmic expansion.

We now turn our attention to the thermodynamical arrow, as determined by the increase of horizon entropy.
The future and past event horizons are defined respectively as\footnote{Note that the definition of $R_h^{-}(t)$ coincides with the {\it particle horizon} used in the cosmology literature. In standard cosmology the lower integration limit corresponds to the initial cosmological singularity, which takes place at a finite value for $t$. However, the model at hand is non-singular and therefore the lower integration limit can be pushed to the infinite past.} \cite{Davies:1988dk,Faraoni_2015}
\be\label{Eq:HorizonsDef}
R_h^{+}(t) = a(t)\int_{t}^{+\infty} \frac{\de t^{\prime}}{a(t^{\prime})}~,\qquad
R_h^{-}(t) = a(t)\int_{-\infty}^{t} \frac{\de t^{\prime}}{a(t^{\prime})}~.
\ee
The presence of a positive cosmological constant in the model at hand ensures that the integrals in Eq.~\eqref{Eq:HorizonsDef} are convergent.
From the definitions \eqref{Eq:HorizonsDef}, we obtain that the horizons obey the following differential equations
\be
\dot{R}_h^{+}(t)=H(t) R_h^{+}(t)-1~,\quad \dot{R}_h^{-}(t)=H(t) R_h^{-}(t)+1~.
\ee
In the regime where $\dot{H}>0$, both horizons are increasing in the expanding phase $H>0$ and decreasing during contraction $H<0$; moreover, they both lie outside the Hubble radius, $\dot{R}_h^{\pm}(t)\geq 1/|H(t)|$. Asymptotically, a de Sitter universe is approached both in the distant past and future, and one has $\lim_{t\to+\infty}R_h^{+}=\lim_{t\to-\infty}R_h^{-}=H_{\infty}^{-1}$. The behaviour of the horizons in the $\dot{H}>0$ regime is consistent with the general analysis in Ref.~\cite{Davies:1988dk}. However, in the proximity of the bounce, where the null energy condition is violated by the effective fluid $\tilde{\rho}+\tilde{p}<0$, such monotonicity properties of the horizons no longer hold. This implies in particular that the minimum value of the horizons is not attained at the bounce at $t=0$.

Given the above, in the following we will limit ourselves to the $\dot{H}>0$ regime. Here the interpretation of the future event horizon as the entropy of the cosmological background is justified \cite{Davies:1988dk}; therefore, the entropy is $\mathcal{S}^{+}(t)\propto (R_h^{+}(t))^2$ and increases with $|t|$, both in the positive and the negative time directions. However, there appears to be a temporal asymmetry, since the entropy $\mathcal{S}^{+}(t)$ is not an even function of time in this model. Naively, this may seem to suggest the existence of a thermodynamical arrow of time for the cosmological background. However, this conclusion is not quite correct, since the apparent time asymmetry merely originates from the definition \eqref{Eq:HorizonsDef} of the event horizon, which is not itself time-symmetric. When time is reversed, also the future and past event horizons must be interchanged. In fact, the entropy of the time-reversed solution (obtained under the transformation $t\to - t$) is given by the area of the {\it past} event horizon $\mathcal{S}^{-}(t)\propto (R_h^{-}(t))^2$. In complete analogy with the previous case, the entropy increases with $|t|$. Furthermore, as a consequence of the symmetry of the background solution \eqref{Eq:ScaleFacSol} under time reversal, we have $R_h^{-}(t)=R_h^{+}(-t)$, which in turn implies $\mathcal{S}^{-}(t)=\mathcal{S}^{+}(-t)$.   Thus, in the model at hand the entropy is indeed invariant under time reversal. Therefore, similarly to the cosmological arrow of time discussed earlier, we conclude that also the thermodynamical arrow of time points in opposite directions in the contracting and expanding branches. If we restrict our attention to the homogeneous and isotropic background (perhaps unsurprisingly) there are no time asymmetries that may enable us to distinguish the two branches. In the next section, we will discuss how time asymmetries do in fact emerge in the evolution of cosmological perturbations.

\section{Time arrow of perturbations in the \texorpdfstring{$\Lambda$CDM}{LCDM} bounce scenario}\label{Sec:3}
We consider the $\Lambda$CDM bounce scenario of Refs.~\cite{Cai:2014jla,Cai:2016hea} and review the evolution of scalar perturbations in their model. As in the simplified model considered in the previous section, also the $\Lambda$CDM bounce is symmetric. The difference between the two models lies in the fact that the $\Lambda$CDM bounce also takes into account the contribution of radiation, which dominates the energy density of the universe at the bounce.
Moreover, there is an intermediate era of radiation domination between the bounce and the matter dominated era.
At late times, as well as in the distant past, radiation becomes sub-leading compared to non-relativistic matter and dark energy, and therefore the evolution of the scale factor in this regime is well approximated by \eqref{Eq:ScaleFacSol}. In both models, general relativity is recovered in the large $|t|$ limit.

The evolution of scalar primordial fluctuations in Fourier space is governed by the Mukhanov-Sasaki equation \cite{Cai:2014jla}
\be\label{Eq:MukhanovSasaki}
v^{\prime\prime}(\eta)+\lf[c_s^2 k^2\lf(1-\frac{2\rho}{\rho_c}\rg)-\frac{z^{\prime\prime}(\eta)}{z(\eta)}\r]v(\eta)=0~,
\ee
where $k$ is the comoving wave number, $z=a \sqrt{\rho+p}/(c_s H)$ (with $c_s$ denoting the speed of sound), and primes denote derivatives with respect to conformal time $\eta=\int\de t/a(t)$. The integration constant in the definition of $\eta$ can be chosen in such a way that $\eta=0$ at the bounce. The Mukhanov-Sasaki variable is related to the comoving curvature perturbation as $v=z\mathcal{R}$.

The equation of motion \eqref{Eq:MukhanovSasaki} is manifestly invariant under time reversal $\eta\to - \eta$. In fact, it follows from the definition that $z$ is odd under time reversal, which in turn implies that the potential $z^{\prime\prime}/z$ is an even function of $\eta$. Therefore, asymmetric solutions for $v(\eta)$ may only originate from initial conditions that explicitly break time-reversal symmetry. In fact, in order to get phenomenologically viable predictions, initial `vacuum' boundary conditions are imposed on the perturbations in the distant past $\eta\to-\infty$, where one assumes the asymptotics $v(\eta)\approx e^{-i c_s k \eta}/\sqrt{2c_s k}$. Modes that are observationally relevant become super-horizon during matter domination in the contracting phase, stay in this regime during radiation domination and throughout the bounce, and eventually re-enter the horizon during the expanding phase. To obtain an almost scale-invariant spectrum, it is important to ensure that observable modes exit the horizon during the matter dominated contraction.
A slight red tilt originates from a small negative contribution to the pressure due to the non-zero cosmological constant. Furthermore, the effective equation of state of the fluid with $\rho=\rho_{\rm m}+\Lambda/(8\pi G)$, $p=-\Lambda/(8\pi G)$ is approximately constant during the $\Lambda$CDM-dominated epoch when observationally relevant modes exit the horizon, i.e.~one has $p=w_{\rm eff}\rho$ with constant $w_{\rm eff}<0$ to a good approximation \cite{Cai:2014jla}.

The evolution of perturbations in this model has been studied using both analytical and numerical methods in Ref.~\cite{Cai:2014jla}, to which we refer the reader for more details. After the bounce and at horizon re-entry, the leading contribution to the curvature perturbation is a constant mode. The scalar power spectrum reads
\be
\Delta_{\cal R}^2=\frac{k^3}{2\pi^2} |{\cal R}|^2\propto \sqrt{\frac{\rho_c}{\rho_{\rm Pl}}} \frac{H_e \ell_{\rm Pl}}{c_s^3}\lf(\frac{8\pi G\rho_c|H_e|}{3 k^4} \rg)^{-3w_{\rm eff}}~,
\ee
where $H_e$ is the Hubble rate at the time of matter-radiation equality in the contracting branch.

The temporal asymmetry in the evolution of perturbations in the model at hand is clear from the numerical simulations \cite{Cai:2014jla}. Moreover, although perturbative modes that were originally in the sub-horizon regime in the distant past will eventually re-enter the horizon in the expanding branch, some of them (i.e.,~those that are relevant for structure formation) will then grow large in amplitude during matter domination, so that linear cosmological perturbation theory is no longer applicable and the evolution becomes nonlinear. This indicates that the temporal asymmetries that already emerged at the linear level due to an asymmetric choice of boundary conditions will eventually lead to a strong asymmetry between the contracting and expanding branches due to the growth of nonlinearities. Therefore, perturbations can determine an arrow of time even in scenarios with a symmetric bounce. Furthermore, perturbations that are both sub-Hubble and super-Jeans during the matter-dominated era in the  contracting branch may become non-linear and, under certain conditions on the speed of sound, lead to the formation of black holes that could propagate through the bounce to the expanding branch \cite{Quintin:2016qro}. The increase in entropy due to black-hole formation determines a thermodynamical arrow of time which agrees with the one defined by the growth of inhomogeneities.

\section{Discussion}\label{Sec:4}
We have seen that a number of different arrows of time can be introduced in bouncing cosmologies.
 At the background level, the cosmological arrow of time always points in the direction of expansion, and therefore away from the bounce. It is also possible to define a thermodynamical arrow of time, pointing in the direction of horizon entropy growth. However, if the bounce is symmetric also the background thermodynamical arrow is unable to introduce any distinction between the two branches. We examined this issue in detail for an exact solution of the effective Friedmann equations of LQC, describing a bouncing cosmology with pressureless dust and a positive cosmological constant. This simple example shows that, in general, the analysis of the evolution of the cosmological background is not sufficient to draw any reliable conclusions about the arrow of time. We have then examined the evolution of scalar perturbations and discussed how the time arrow of inhomogeneities growth originates from initial conditions that break time-reversal symmetry. A specific choice of (vacuum) initial conditions is made in the distant past, which amounts to a Past Hypothesis. This plays an analogous role to Penrose's Weyl hypothesis in standard cosmology in selecting a special initial state for the universe.
 
Further arrows of time can be introduced in the case of asymmetric bounces, which we have not discussed in this contribution. For instance, the bounce may be preceded by an epoch of ultra-slow contraction, as in ekpyrotic models, or there could be an epoch of inflation following the bounce. In both cases, such early universe scenarios single out a preferred time direction that discriminates the past from the future. This is best understood by introducing some deviations from the perfect flatness and isotropy of the FLRW cosmological background.
Both in the cases of ekpyrotic and inflationary models there is only one direction of time evolution that leads to the suppression of shear anisotropies and spatial curvature starting from generic initial conditions for the homogeneous background, as ensured by the cosmic no-hair theorems \cite{Barrow:1987ia,Erickson:2003zm}.
More precisely, the arrow of time points in the direction of ekpyrotic contraction and inflationary expansion---a reversed time arrow would be incompatible with the observed properties of our universe.

We stress that the issue of determining an arrow of time is common to all early universe scenarios---not just bouncing cosmologies. In this contribution, the dynamics of the cosmological background and perturbations in bouncing models has been studied at a classical effective level, focusing on a generalization of the matter bounce scenario in LQC. We have seen that in this scenario the arrow of time follows from a specific choice of initial conditions for cosmological perturbations, which are imposed in the distant past in the contracting branch and serve as an example of the Past Hypothesis.
However, in order to fully address the arrow of time problem one should work in a full theory of quantum gravity, for which we lack a complete formulation at present. To complicate matters further, in quantum gravity the time arrow problem is also closely connected with the {\it problem of time}, i.e.~the absence of time at a fundamental level~\cite{KUCHA__2011}. 

Lastly, if spacetime emerges dynamically from a pre-geometric structure, as argued for instance within the context of GFT and in the {\it causal sets} approach to quantum gravity (see contributions by Oriti \cite{Oriti:2024elx} and Forgione \cite{Forgione:2024upw} in this volume, and Ref.~\cite{DeBianchi:2023yys}), the problem of defining an `initial state' for the universe may require a reframing that makes no reference to the emergent spacetime geometry. Within this framework, it would be interesting to investigate whether the initial conditions of the universe could be explained from first principles, for instance by means of a dynamical mechanism, or if a Past Hypothesis (i.e.,~a law-like prescription similar to the Weyl curvature hypothesis) is still required.

\section*{Acknowledgments}

I am grateful to Silvia De Bianchi for the invitation to contribute to the volume ``Time and Timelessness in Fundamental Physics and Cosmology'', and would like to thank Edward Wilson-Ewing and Sergio Zerbini for helpful comments. The author acknowledges support from INFN, iniziative specifiche QUAGRAP and GeoSymQFT.

\bibliography{arrow_refs}

\begin{thebibliography}{78}%
\makeatletter
\providecommand \@ifxundefined [1]{%
 \@ifx{#1\undefined}
}%
\providecommand \@ifnum [1]{%
 \ifnum #1\expandafter \@firstoftwo
 \else \expandafter \@secondoftwo
 \fi
}%
\providecommand \@ifx [1]{%
 \ifx #1\expandafter \@firstoftwo
 \else \expandafter \@secondoftwo
 \fi
}%
\providecommand \natexlab [1]{#1}%
\providecommand \enquote  [1]{``#1''}%
\providecommand \bibnamefont  [1]{#1}%
\providecommand \bibfnamefont [1]{#1}%
\providecommand \citenamefont [1]{#1}%
\providecommand \href@noop [0]{\@secondoftwo}%
\providecommand \href [0]{\begingroup \@sanitize@url \@href}%
\providecommand \@href[1]{\@@startlink{#1}\@@href}%
\providecommand \@@href[1]{\endgroup#1\@@endlink}%
\providecommand \@sanitize@url [0]{\catcode `\\12\catcode `\$12\catcode
  `\&12\catcode `\#12\catcode `\^12\catcode `\_12\catcode `\%12\relax}%
\providecommand \@@startlink[1]{}%
\providecommand \@@endlink[0]{}%
\providecommand \url  [0]{\begingroup\@sanitize@url \@url }%
\providecommand \@url [1]{\endgroup\@href {#1}{\urlprefix }}%
\providecommand \urlprefix  [0]{URL }%
\providecommand \Eprint [0]{\href }%
\providecommand \doibase [0]{http://dx.doi.org/}%
\providecommand \selectlanguage [0]{\@gobble}%
\providecommand \bibinfo  [0]{\@secondoftwo}%
\providecommand \bibfield  [0]{\@secondoftwo}%
\providecommand \translation [1]{[#1]}%
\providecommand \BibitemOpen [0]{}%
\providecommand \bibitemStop [0]{}%
\providecommand \bibitemNoStop [0]{.\EOS\space}%
\providecommand \EOS [0]{\spacefactor3000\relax}%
\providecommand \BibitemShut  [1]{\csname bibitem#1\endcsname}%
\let\auto@bib@innerbib\@empty
\bibitem [{\citenamefont {Hawking}(1985)}]{Hawking:1985af}%
  \BibitemOpen
  \bibfield  {author} {\bibinfo {author} {\bibfnamefont {S.~W.}\ \bibnamefont
  {Hawking}},\ }\bibfield  {title} {\enquote {\bibinfo {title} {{The Arrow of
  Time in Cosmology}},}\ }\href {\doibase 10.1103/PhysRevD.32.2489} {\bibfield
  {journal} {\bibinfo  {journal} {Phys. Rev. D}\ }\textbf {\bibinfo {volume}
  {32}},\ \bibinfo {pages} {2489} (\bibinfo {year} {1985})}\BibitemShut
  {NoStop}%
\bibitem [{\citenamefont {Gell-Mann}\ and\ \citenamefont
  {Hartle}(1991)}]{Gell-Mann:1991kdm}%
  \BibitemOpen
  \bibfield  {author} {\bibinfo {author} {\bibfnamefont {Murray}\ \bibnamefont
  {Gell-Mann}}\ and\ \bibinfo {author} {\bibfnamefont {James~B.}\ \bibnamefont
  {Hartle}},\ }\bibfield  {title} {\enquote {\bibinfo {title} {{Time symmetry
  and asymmetry in quantum mechanics and quantum cosmology}},}\ }in\ \href@noop
  {} {\emph {\bibinfo {booktitle} {{The First International A. D. Sakharov
  Conference on Physics}}}}\ (\bibinfo {year} {1991})\ \Eprint
  {http://arxiv.org/abs/gr-qc/9304023} {arXiv:gr-qc/9304023} \BibitemShut
  {NoStop}%
\bibitem [{\citenamefont {Ellis}(2013)}]{ELLIS2013242}%
  \BibitemOpen
  \bibfield  {author} {\bibinfo {author} {\bibfnamefont {George
  Francis~Rayner}\ \bibnamefont {Ellis}},\ }\bibfield  {title} {\enquote
  {\bibinfo {title} {The arrow of time and the nature of spacetime},}\ }\href
  {\doibase https://doi.org/10.1016/j.shpsb.2013.06.002} {\bibfield  {journal}
  {\bibinfo  {journal} {Studies in History and Philosophy of Science Part B:
  Studies in History and Philosophy of Modern Physics}\ }\textbf {\bibinfo
  {volume} {44}},\ \bibinfo {pages} {242--262} (\bibinfo {year}
  {2013})}\BibitemShut {NoStop}%
\bibitem [{\citenamefont {Zeh}(2011)}]{Zeh_2011}%
  \BibitemOpen
  \bibfield  {author} {\bibinfo {author} {\bibfnamefont {H.~Dieter}\
  \bibnamefont {Zeh}},\ }\enquote {\bibinfo {title} {Open questions regarding
  the arrow of time},}\ in\ \href {\doibase 10.1007/978-3-642-23259-6_11}
  {\emph {\bibinfo {booktitle} {The Arrows of Time}}}\ (\bibinfo  {publisher}
  {Springer Berlin Heidelberg},\ \bibinfo {year} {2011})\ pp.\ \bibinfo {pages}
  {205--217}\BibitemShut {NoStop}%
\bibitem [{\citenamefont {Forgione}(2022)}]{Forgione:2022nms}%
  \BibitemOpen
  \bibfield  {author} {\bibinfo {author} {\bibfnamefont {Marco}\ \bibnamefont
  {Forgione}},\ }\emph {\bibinfo {title} {{History and Philosophy of
  Feynman\textquoteright{}s Electrodynamics: From the Absorber Theory of
  Radiation to Feynman Diagrams}}},\ \href@noop {} {Ph.D. thesis},\ \bibinfo
  {school} {University of South Carolina} (\bibinfo {year} {2022})\BibitemShut
  {NoStop}%
\bibitem [{\citenamefont {Schulman}(1999)}]{Schulman:1999kg}%
  \BibitemOpen
  \bibfield  {author} {\bibinfo {author} {\bibfnamefont {L.~S.}\ \bibnamefont
  {Schulman}},\ }\bibfield  {title} {\enquote {\bibinfo {title} {{Opposite
  thermodynamic arrows of time}},}\ }\href {\doibase
  10.1103/PhysRevLett.83.5419} {\bibfield  {journal} {\bibinfo  {journal}
  {Phys. Rev. Lett.}\ }\textbf {\bibinfo {volume} {83}},\ \bibinfo {pages}
  {5419--5422} (\bibinfo {year} {1999})},\ \Eprint
  {http://arxiv.org/abs/cond-mat/9911101} {arXiv:cond-mat/9911101} \BibitemShut
  {NoStop}%
\bibitem [{\citenamefont {Zeh}(2005)}]{Zeh_2005}%
  \BibitemOpen
  \bibfield  {author} {\bibinfo {author} {\bibfnamefont {H.}~\bibnamefont
  {Zeh}},\ }\bibfield  {title} {\enquote {\bibinfo {title} {Remarks on the
  compatibility of opposite arrows of time},}\ }\href {\doibase
  10.3390/e7040199} {\bibfield  {journal} {\bibinfo  {journal} {Entropy}\
  }\textbf {\bibinfo {volume} {7}},\ \bibinfo {pages} {199--207} (\bibinfo
  {year} {2005})}\BibitemShut {NoStop}%
\bibitem [{\citenamefont {Kiefer}(2011)}]{Kiefer_2011}%
  \BibitemOpen
  \bibfield  {author} {\bibinfo {author} {\bibfnamefont {Claus}\ \bibnamefont
  {Kiefer}},\ }\enquote {\bibinfo {title} {Can the arrow of time be understood
  from quantum cosmology?}}\ in\ \href {\doibase 10.1007/978-3-642-23259-6_10}
  {\emph {\bibinfo {booktitle} {The Arrows of Time}}}\ (\bibinfo  {publisher}
  {Springer Berlin Heidelberg},\ \bibinfo {year} {2011})\ pp.\ \bibinfo {pages}
  {191--203}\BibitemShut {NoStop}%
\bibitem [{\citenamefont {Gold}(1962)}]{Gold_1962}%
  \BibitemOpen
  \bibfield  {author} {\bibinfo {author} {\bibfnamefont {T.}~\bibnamefont
  {Gold}},\ }\bibfield  {title} {\enquote {\bibinfo {title} {The arrow of
  time},}\ }\href {\doibase 10.1119/1.1942052} {\bibfield  {journal} {\bibinfo
  {journal} {American Journal of Physics}\ }\textbf {\bibinfo {volume} {30}},\
  \bibinfo {pages} {403--410} (\bibinfo {year} {1962})}\BibitemShut {NoStop}%
\bibitem [{\citenamefont {Penrose}(1979)}]{Penrose:1979azm}%
  \BibitemOpen
  \bibfield  {author} {\bibinfo {author} {\bibfnamefont {R.}~\bibnamefont
  {Penrose}},\ }\bibfield  {title} {\enquote {\bibinfo {title} {{Singularities
  and time-asymmetry}},}\ \ }(\bibinfo {year} {1979})\ pp.\ \bibinfo {pages}
  {581--638}\BibitemShut {NoStop}%
\bibitem [{\citenamefont {Zeldovich}\ and\ \citenamefont
  {Novikov}(1983)}]{Zeldovich:1983cr}%
  \BibitemOpen
  \bibfield  {author} {\bibinfo {author} {\bibfnamefont {Ya.~B.}\ \bibnamefont
  {Zeldovich}}\ and\ \bibinfo {author} {\bibfnamefont {I.~D.}\ \bibnamefont
  {Novikov}},\ }\href@noop {} {\emph {\bibinfo {title} {{Relativistic
  astrophysics. Vol. 2. The structure and evolution of the Universe}}}}\
  (\bibinfo  {publisher} {University of Chicago Press},\ \bibinfo {year}
  {1983})\BibitemShut {NoStop}%
\bibitem [{\citenamefont {Page}(1985)}]{Page:1985ei}%
  \BibitemOpen
  \bibfield  {author} {\bibinfo {author} {\bibfnamefont {Don~N.}\ \bibnamefont
  {Page}},\ }\bibfield  {title} {\enquote {\bibinfo {title} {{Will Entropy
  Decrease if the Universe Recollapses?}}}\ }\href {\doibase
  10.1103/PhysRevD.32.2496} {\bibfield  {journal} {\bibinfo  {journal} {Phys.
  Rev. D}\ }\textbf {\bibinfo {volume} {32}},\ \bibinfo {pages} {2496}
  (\bibinfo {year} {1985})}\BibitemShut {NoStop}%
\bibitem [{\citenamefont {Kiefer}\ and\ \citenamefont
  {Zeh}(1995)}]{Kiefer:1994gp}%
  \BibitemOpen
  \bibfield  {author} {\bibinfo {author} {\bibfnamefont {C.}~\bibnamefont
  {Kiefer}}\ and\ \bibinfo {author} {\bibfnamefont {H.~D.}\ \bibnamefont
  {Zeh}},\ }\bibfield  {title} {\enquote {\bibinfo {title} {{Arrow of time in a
  recollapsing quantum universe}},}\ }\href {\doibase 10.1103/PhysRevD.51.4145}
  {\bibfield  {journal} {\bibinfo  {journal} {Phys. Rev. D}\ }\textbf {\bibinfo
  {volume} {51}},\ \bibinfo {pages} {4145--4153} (\bibinfo {year} {1995})},\
  \Eprint {http://arxiv.org/abs/gr-qc/9402036} {arXiv:gr-qc/9402036}
  \BibitemShut {NoStop}%
\bibitem [{\citenamefont {Penrose}(1989)}]{Penrose:1988mg}%
  \BibitemOpen
  \bibfield  {author} {\bibinfo {author} {\bibfnamefont {R.}~\bibnamefont
  {Penrose}},\ }\bibfield  {title} {\enquote {\bibinfo {title} {{Difficulties
  with inflationary cosmology}},}\ }\href {\doibase
  10.1111/j.1749-6632.1989.tb50513.x} {\bibfield  {journal} {\bibinfo
  {journal} {Annals N. Y. Acad. Sci.}\ }\textbf {\bibinfo {volume} {571}},\
  \bibinfo {pages} {249--264} (\bibinfo {year} {1989})}\BibitemShut {NoStop}%
\bibitem [{\citenamefont {Penrose}(2006)}]{Penrose:2006zz}%
  \BibitemOpen
  \bibfield  {author} {\bibinfo {author} {\bibfnamefont {R.}~\bibnamefont
  {Penrose}},\ }\bibfield  {title} {\enquote {\bibinfo {title} {{Before the big
  bang: An outrageous new perspective and its implications for particle
  physics}},}\ }\href@noop {} {\bibfield  {journal} {\bibinfo  {journal} {Conf.
  Proc. C}\ }\textbf {\bibinfo {volume} {060626}},\ \bibinfo {pages}
  {2759--2767} (\bibinfo {year} {2006})}\BibitemShut {NoStop}%
\bibitem [{\citenamefont {Bardeen}\ \emph {et~al.}(1973)\citenamefont
  {Bardeen}, \citenamefont {Carter},\ and\ \citenamefont
  {Hawking}}]{Bardeen:1973gs}%
  \BibitemOpen
  \bibfield  {author} {\bibinfo {author} {\bibfnamefont {James~M.}\
  \bibnamefont {Bardeen}}, \bibinfo {author} {\bibfnamefont {B.}~\bibnamefont
  {Carter}}, \ and\ \bibinfo {author} {\bibfnamefont {S.~W.}\ \bibnamefont
  {Hawking}},\ }\bibfield  {title} {\enquote {\bibinfo {title} {{The Four laws
  of black hole mechanics}},}\ }\href {\doibase 10.1007/BF01645742} {\bibfield
  {journal} {\bibinfo  {journal} {Commun. Math. Phys.}\ }\textbf {\bibinfo
  {volume} {31}},\ \bibinfo {pages} {161--170} (\bibinfo {year}
  {1973})}\BibitemShut {NoStop}%
\bibitem [{\citenamefont {Davies}(1988)}]{Davies:1988dk}%
  \BibitemOpen
  \bibfield  {author} {\bibinfo {author} {\bibfnamefont {P.~C.~W.}\
  \bibnamefont {Davies}},\ }\bibfield  {title} {\enquote {\bibinfo {title}
  {{Cosmological Horizons and Entropy}},}\ }\href {\doibase
  10.1088/0264-9381/5/10/013} {\bibfield  {journal} {\bibinfo  {journal}
  {Class. Quant. Grav.}\ }\textbf {\bibinfo {volume} {5}},\ \bibinfo {pages}
  {1349} (\bibinfo {year} {1988})}\BibitemShut {NoStop}%
\bibitem [{\citenamefont {Jacobson}\ and\ \citenamefont
  {Parentani}(2003)}]{Jacobson:2003wv}%
  \BibitemOpen
  \bibfield  {author} {\bibinfo {author} {\bibfnamefont {Ted}\ \bibnamefont
  {Jacobson}}\ and\ \bibinfo {author} {\bibfnamefont {Renaud}\ \bibnamefont
  {Parentani}},\ }\bibfield  {title} {\enquote {\bibinfo {title} {{Horizon
  entropy}},}\ }\href {\doibase 10.1023/A:1023785123428} {\bibfield  {journal}
  {\bibinfo  {journal} {Found. Phys.}\ }\textbf {\bibinfo {volume} {33}},\
  \bibinfo {pages} {323--348} (\bibinfo {year} {2003})},\ \Eprint
  {http://arxiv.org/abs/gr-qc/0302099} {arXiv:gr-qc/0302099} \BibitemShut
  {NoStop}%
\bibitem [{\citenamefont {Husain}(1988)}]{Husain:1988jf}%
  \BibitemOpen
  \bibfield  {author} {\bibinfo {author} {\bibfnamefont {Viqar}\ \bibnamefont
  {Husain}},\ }\bibfield  {title} {\enquote {\bibinfo {title} {{The Weyl Tensor
  and Gravitational Entropy}},}\ }\href {\doibase 10.1103/PhysRevD.38.3314}
  {\bibfield  {journal} {\bibinfo  {journal} {Phys. Rev. D}\ }\textbf {\bibinfo
  {volume} {38}},\ \bibinfo {pages} {3314--3317} (\bibinfo {year}
  {1988})}\BibitemShut {NoStop}%
\bibitem [{\citenamefont {Bonnor}(1987)}]{BONNOR1987305}%
  \BibitemOpen
  \bibfield  {author} {\bibinfo {author} {\bibfnamefont {W.B.}\ \bibnamefont
  {Bonnor}},\ }\bibfield  {title} {\enquote {\bibinfo {title} {Arrow of time
  for a collapsing, radiating sphere},}\ }\href {\doibase
  https://doi.org/10.1016/0375-9601(87)90830-9} {\bibfield  {journal} {\bibinfo
   {journal} {Physics Letters A}\ }\textbf {\bibinfo {volume} {122}},\ \bibinfo
  {pages} {305--308} (\bibinfo {year} {1987})}\BibitemShut {NoStop}%
\bibitem [{\citenamefont {Goode}\ \emph {et~al.}(1992)\citenamefont {Goode},
  \citenamefont {Coley},\ and\ \citenamefont {Wainwright}}]{Goode:1992pp}%
  \BibitemOpen
  \bibfield  {author} {\bibinfo {author} {\bibfnamefont {S.~W.}\ \bibnamefont
  {Goode}}, \bibinfo {author} {\bibfnamefont {A.~A.}\ \bibnamefont {Coley}}, \
  and\ \bibinfo {author} {\bibfnamefont {J.}~\bibnamefont {Wainwright}},\
  }\bibfield  {title} {\enquote {\bibinfo {title} {{The Isotropic singularity
  in cosmology}},}\ }\href {\doibase 10.1088/0264-9381/9/2/010} {\bibfield
  {journal} {\bibinfo  {journal} {Class. Quant. Grav.}\ }\textbf {\bibinfo
  {volume} {9}},\ \bibinfo {pages} {445--455} (\bibinfo {year}
  {1992})}\BibitemShut {NoStop}%
\bibitem [{\citenamefont {Clifton}\ \emph {et~al.}(2013)\citenamefont
  {Clifton}, \citenamefont {Ellis},\ and\ \citenamefont
  {Tavakol}}]{Clifton:2013dha}%
  \BibitemOpen
  \bibfield  {author} {\bibinfo {author} {\bibfnamefont {Timothy}\ \bibnamefont
  {Clifton}}, \bibinfo {author} {\bibfnamefont {George F.~R.}\ \bibnamefont
  {Ellis}}, \ and\ \bibinfo {author} {\bibfnamefont {Reza}\ \bibnamefont
  {Tavakol}},\ }\bibfield  {title} {\enquote {\bibinfo {title} {{A
  Gravitational Entropy Proposal}},}\ }\href {\doibase
  10.1088/0264-9381/30/12/125009} {\bibfield  {journal} {\bibinfo  {journal}
  {Class. Quant. Grav.}\ }\textbf {\bibinfo {volume} {30}},\ \bibinfo {pages}
  {125009} (\bibinfo {year} {2013})},\ \Eprint {http://arxiv.org/abs/1303.5612}
  {arXiv:1303.5612 [gr-qc]} \BibitemShut {NoStop}%
\bibitem [{\citenamefont {Penrose}(1994)}]{Penrose:1994de}%
  \BibitemOpen
  \bibfield  {author} {\bibinfo {author} {\bibfnamefont {Roger}\ \bibnamefont
  {Penrose}},\ }\bibfield  {title} {\enquote {\bibinfo {title} {{On the Second
  law of thermodynamics}},}\ }\href {\doibase 10.1007/BF02186840} {\bibfield
  {journal} {\bibinfo  {journal} {J. Statist. Phys.}\ }\textbf {\bibinfo
  {volume} {77}},\ \bibinfo {pages} {217--221} (\bibinfo {year}
  {1994})}\BibitemShut {NoStop}%
\bibitem [{\citenamefont {Belinsky}\ \emph {et~al.}(1982)\citenamefont
  {Belinsky}, \citenamefont {Khalatnikov},\ and\ \citenamefont
  {Lifshitz}}]{Belinsky:1982pk}%
  \BibitemOpen
  \bibfield  {author} {\bibinfo {author} {\bibfnamefont {V.~a.}\ \bibnamefont
  {Belinsky}}, \bibinfo {author} {\bibfnamefont {I.~m.}\ \bibnamefont
  {Khalatnikov}}, \ and\ \bibinfo {author} {\bibfnamefont {E.~m.}\ \bibnamefont
  {Lifshitz}},\ }\bibfield  {title} {\enquote {\bibinfo {title} {{A General
  Solution of the Einstein Equations with a Time Singularity}},}\ }\href
  {\doibase 10.1080/00018738200101428} {\bibfield  {journal} {\bibinfo
  {journal} {Adv. Phys.}\ }\textbf {\bibinfo {volume} {31}},\ \bibinfo {pages}
  {639--667} (\bibinfo {year} {1982})}\BibitemShut {NoStop}%
\bibitem [{\citenamefont {Albert}(2000)}]{Albert2000-ALBTAC-4}%
  \BibitemOpen
  \bibfield  {author} {\bibinfo {author} {\bibfnamefont {David~Z.}\
  \bibnamefont {Albert}},\ }\href@noop {} {\emph {\bibinfo {title} {Time and
  Chance}}}\ (\bibinfo  {publisher} {Harvard University Press},\ \bibinfo
  {address} {Cambridge, Mass.},\ \bibinfo {year} {2000})\BibitemShut {NoStop}%
\bibitem [{\citenamefont {Schiffrin}\ and\ \citenamefont
  {Wald}(2012)}]{Schiffrin:2012zf}%
  \BibitemOpen
  \bibfield  {author} {\bibinfo {author} {\bibfnamefont {Joshua~S.}\
  \bibnamefont {Schiffrin}}\ and\ \bibinfo {author} {\bibfnamefont {Robert~M.}\
  \bibnamefont {Wald}},\ }\bibfield  {title} {\enquote {\bibinfo {title}
  {{Measure and Probability in Cosmology}},}\ }\href {\doibase
  10.1103/PhysRevD.86.023521} {\bibfield  {journal} {\bibinfo  {journal} {Phys.
  Rev. D}\ }\textbf {\bibinfo {volume} {86}},\ \bibinfo {pages} {023521}
  (\bibinfo {year} {2012})},\ \Eprint {http://arxiv.org/abs/1202.1818}
  {arXiv:1202.1818 [gr-qc]} \BibitemShut {NoStop}%
\bibitem [{\citenamefont {Curiel}(2015)}]{Curiel:2015oea}%
  \BibitemOpen
  \bibfield  {author} {\bibinfo {author} {\bibfnamefont {Erik}\ \bibnamefont
  {Curiel}},\ }\bibfield  {title} {\enquote {\bibinfo {title} {{Measure,
  Topology and Probabilistic Reasoning in Cosmology}},}\ }\href@noop {} {\
  (\bibinfo {year} {2015})},\ \Eprint {http://arxiv.org/abs/1509.01878}
  {arXiv:1509.01878 [gr-qc]} \BibitemShut {NoStop}%
\bibitem [{\citenamefont {Price}(2004)}]{Price2004-PRIOTO-2}%
  \BibitemOpen
  \bibfield  {author} {\bibinfo {author} {\bibfnamefont {Huw}\ \bibnamefont
  {Price}},\ }\bibfield  {title} {\enquote {\bibinfo {title} {On the origins of
  the arrow of time: Why there is still a puzzle about the low entropy past},}\
  }in\ \href@noop {} {\emph {\bibinfo {booktitle} {Contemporary Debates in
  Philosophy of Science}}},\ \bibinfo {editor} {edited by\ \bibinfo {editor}
  {\bibfnamefont {Christopher}\ \bibnamefont {Hitchcock}}}\ (\bibinfo
  {publisher} {Blackwell},\ \bibinfo {year} {2004})\ pp.\ \bibinfo {pages}
  {219--239}\BibitemShut {NoStop}%
\bibitem [{\citenamefont {Callender}(2004)}]{Callender2004-CALTIN-2}%
  \BibitemOpen
  \bibfield  {author} {\bibinfo {author} {\bibfnamefont {Craig}\ \bibnamefont
  {Callender}},\ }\bibfield  {title} {\enquote {\bibinfo {title} {There is no
  puzzle about the low entropy past},}\ }in\ \href@noop {} {\emph {\bibinfo
  {booktitle} {Contemporary Debates in Philosophy of Science}}},\ \bibinfo
  {editor} {edited by\ \bibinfo {editor} {\bibfnamefont {Christopher}\
  \bibnamefont {Hitchcock}}}\ (\bibinfo  {publisher} {Blackwell},\ \bibinfo
  {year} {2004})\ pp.\ \bibinfo {pages} {240--255}\BibitemShut {NoStop}%
\bibitem [{\citenamefont {Earman}(2006)}]{Earman2006-EARTPH-3}%
  \BibitemOpen
  \bibfield  {author} {\bibinfo {author} {\bibfnamefont {John}\ \bibnamefont
  {Earman}},\ }\bibfield  {title} {\enquote {\bibinfo {title} {The ``past
  hypothesis'': Not even false},}\ }\href {\doibase
  10.1016/j.shpsb.2006.03.002} {\bibfield  {journal} {\bibinfo  {journal}
  {Studies in History and Philosophy of Modern Physics}\ }\textbf {\bibinfo
  {volume} {37}},\ \bibinfo {pages} {399--430} (\bibinfo {year}
  {2006})}\BibitemShut {NoStop}%
\bibitem [{\citenamefont {Gryb}(2021)}]{Gryb:2020thg}%
  \BibitemOpen
  \bibfield  {author} {\bibinfo {author} {\bibfnamefont {Sean}\ \bibnamefont
  {Gryb}},\ }\bibfield  {title} {\enquote {\bibinfo {title} {{New Difficulties
  for the Past Hypothesis}},}\ }\href {\doibase 10.1086/712879} {\bibfield
  {journal} {\bibinfo  {journal} {Phil. Sci.}\ }\textbf {\bibinfo {volume}
  {88}},\ \bibinfo {pages} {511--532} (\bibinfo {year} {2021})},\ \Eprint
  {http://arxiv.org/abs/2006.01576} {arXiv:2006.01576 [physics.hist-ph]}
  \BibitemShut {NoStop}%
\bibitem [{\citenamefont {Wald}(2006)}]{Wald:2005cb}%
  \BibitemOpen
  \bibfield  {author} {\bibinfo {author} {\bibfnamefont {Robert~M.}\
  \bibnamefont {Wald}},\ }\bibfield  {title} {\enquote {\bibinfo {title} {{The
  Arrow of time and the initial conditions of the universe}},}\ }\href
  {\doibase 10.1016/j.shpsb.2006.03.005} {\bibfield  {journal} {\bibinfo
  {journal} {Stud. Hist. Phil. Mod. Phys.}\ }\textbf {\bibinfo {volume} {37}},\
  \bibinfo {pages} {394--398} (\bibinfo {year} {2006})},\ \Eprint
  {http://arxiv.org/abs/gr-qc/0507094} {arXiv:gr-qc/0507094} \BibitemShut
  {NoStop}%
\bibitem [{\citenamefont {Davies}(2014)}]{Davies:2014}%
  \BibitemOpen
  \bibfield  {author} {\bibinfo {author} {\bibfnamefont {P.C.W.}\ \bibnamefont
  {Davies}},\ }\bibfield  {title} {\enquote {\bibinfo {title} {The arrow of
  time},}\ }\href@noop {} {\bibfield  {journal} {\bibinfo  {journal} {Euresis}\
  }\textbf {\bibinfo {volume} {7}},\ \bibinfo {pages} {25--37} (\bibinfo {year}
  {2014})}\BibitemShut {NoStop}%
\bibitem [{\citenamefont {Vilenkin}(1983)}]{Vilenkin:1983xq}%
  \BibitemOpen
  \bibfield  {author} {\bibinfo {author} {\bibfnamefont {Alexander}\
  \bibnamefont {Vilenkin}},\ }\bibfield  {title} {\enquote {\bibinfo {title}
  {{The Birth of Inflationary Universes}},}\ }\href {\doibase
  10.1103/PhysRevD.27.2848} {\bibfield  {journal} {\bibinfo  {journal} {Phys.
  Rev. D}\ }\textbf {\bibinfo {volume} {27}},\ \bibinfo {pages} {2848}
  (\bibinfo {year} {1983})}\BibitemShut {NoStop}%
\bibitem [{\citenamefont {Hartle}\ and\ \citenamefont
  {Hawking}(1983)}]{Hartle:1983ai}%
  \BibitemOpen
  \bibfield  {author} {\bibinfo {author} {\bibfnamefont {J.~B.}\ \bibnamefont
  {Hartle}}\ and\ \bibinfo {author} {\bibfnamefont {S.~W.}\ \bibnamefont
  {Hawking}},\ }\bibfield  {title} {\enquote {\bibinfo {title} {{Wave Function
  of the Universe}},}\ }\href {\doibase 10.1103/PhysRevD.28.2960} {\bibfield
  {journal} {\bibinfo  {journal} {Phys. Rev. D}\ }\textbf {\bibinfo {volume}
  {28}},\ \bibinfo {pages} {2960--2975} (\bibinfo {year} {1983})}\BibitemShut
  {NoStop}%
\bibitem [{\citenamefont {Feldbrugge}\ \emph {et~al.}(2017)\citenamefont
  {Feldbrugge}, \citenamefont {Lehners},\ and\ \citenamefont
  {Turok}}]{Feldbrugge:2017fcc}%
  \BibitemOpen
  \bibfield  {author} {\bibinfo {author} {\bibfnamefont {Job}\ \bibnamefont
  {Feldbrugge}}, \bibinfo {author} {\bibfnamefont {Jean-Luc}\ \bibnamefont
  {Lehners}}, \ and\ \bibinfo {author} {\bibfnamefont {Neil}\ \bibnamefont
  {Turok}},\ }\bibfield  {title} {\enquote {\bibinfo {title} {{No smooth
  beginning for spacetime}},}\ }\href {\doibase 10.1103/PhysRevLett.119.171301}
  {\bibfield  {journal} {\bibinfo  {journal} {Phys. Rev. Lett.}\ }\textbf
  {\bibinfo {volume} {119}},\ \bibinfo {pages} {171301} (\bibinfo {year}
  {2017})},\ \Eprint {http://arxiv.org/abs/1705.00192} {arXiv:1705.00192
  [hep-th]} \BibitemShut {NoStop}%
\bibitem [{\citenamefont {Aguirre}\ and\ \citenamefont
  {Gratton}(2003)}]{Aguirre:2003ck}%
  \BibitemOpen
  \bibfield  {author} {\bibinfo {author} {\bibfnamefont {Anthony}\ \bibnamefont
  {Aguirre}}\ and\ \bibinfo {author} {\bibfnamefont {Steven}\ \bibnamefont
  {Gratton}},\ }\bibfield  {title} {\enquote {\bibinfo {title} {{Inflation
  without a beginning: A Null boundary proposal}},}\ }\href {\doibase
  10.1103/PhysRevD.67.083515} {\bibfield  {journal} {\bibinfo  {journal} {Phys.
  Rev. D}\ }\textbf {\bibinfo {volume} {67}},\ \bibinfo {pages} {083515}
  (\bibinfo {year} {2003})},\ \Eprint {http://arxiv.org/abs/gr-qc/0301042}
  {arXiv:gr-qc/0301042} \BibitemShut {NoStop}%
\bibitem [{\citenamefont {Carroll}\ and\ \citenamefont
  {Chen}(2004)}]{Carroll:2004pn}%
  \BibitemOpen
  \bibfield  {author} {\bibinfo {author} {\bibfnamefont {Sean~M.}\ \bibnamefont
  {Carroll}}\ and\ \bibinfo {author} {\bibfnamefont {Jennifer}\ \bibnamefont
  {Chen}},\ }\bibfield  {title} {\enquote {\bibinfo {title} {{Spontaneous
  inflation and the origin of the arrow of time}},}\ }\href@noop {} {\
  (\bibinfo {year} {2004})},\ \Eprint {http://arxiv.org/abs/hep-th/0410270}
  {arXiv:hep-th/0410270} \BibitemShut {NoStop}%
\bibitem [{\citenamefont {Vilenkin}(2013)}]{Vilenkin:2013rza}%
  \BibitemOpen
  \bibfield  {author} {\bibinfo {author} {\bibfnamefont {Alexander}\
  \bibnamefont {Vilenkin}},\ }\bibfield  {title} {\enquote {\bibinfo {title}
  {{Arrows of time and the beginning of the universe}},}\ }\href {\doibase
  10.1103/PhysRevD.88.043516} {\bibfield  {journal} {\bibinfo  {journal} {Phys.
  Rev. D}\ }\textbf {\bibinfo {volume} {88}},\ \bibinfo {pages} {043516}
  (\bibinfo {year} {2013})},\ \Eprint {http://arxiv.org/abs/1305.3836}
  {arXiv:1305.3836 [hep-th]} \BibitemShut {NoStop}%
\bibitem [{\citenamefont {Barbour}\ \emph {et~al.}(2014)\citenamefont
  {Barbour}, \citenamefont {Koslowski},\ and\ \citenamefont
  {Mercati}}]{Barbour:2014bga}%
  \BibitemOpen
  \bibfield  {author} {\bibinfo {author} {\bibfnamefont {Julian}\ \bibnamefont
  {Barbour}}, \bibinfo {author} {\bibfnamefont {Tim}\ \bibnamefont
  {Koslowski}}, \ and\ \bibinfo {author} {\bibfnamefont {Flavio}\ \bibnamefont
  {Mercati}},\ }\bibfield  {title} {\enquote {\bibinfo {title} {{Identification
  of a gravitational arrow of time}},}\ }\href {\doibase
  10.1103/PhysRevLett.113.181101} {\bibfield  {journal} {\bibinfo  {journal}
  {Phys. Rev. Lett.}\ }\textbf {\bibinfo {volume} {113}},\ \bibinfo {pages}
  {181101} (\bibinfo {year} {2014})},\ \Eprint {http://arxiv.org/abs/1409.0917}
  {arXiv:1409.0917 [gr-qc]} \BibitemShut {NoStop}%
\bibitem [{\citenamefont {Deutsch}\ and\ \citenamefont
  {Aguirre}(2022)}]{Deutsch:2021ylr}%
  \BibitemOpen
  \bibfield  {author} {\bibinfo {author} {\bibfnamefont {J.~M.}\ \bibnamefont
  {Deutsch}}\ and\ \bibinfo {author} {\bibfnamefont {Anthony}\ \bibnamefont
  {Aguirre}},\ }\bibfield  {title} {\enquote {\bibinfo {title} {{State-to-State
  Cosmology: A New View on the Cosmological Arrow of Time and the Past
  Hypothesis}},}\ }\href {\doibase 10.1007/s10701-022-00597-3} {\bibfield
  {journal} {\bibinfo  {journal} {Found. Phys.}\ }\textbf {\bibinfo {volume}
  {52}},\ \bibinfo {pages} {82} (\bibinfo {year} {2022})},\ \Eprint
  {http://arxiv.org/abs/2106.15692} {arXiv:2106.15692 [physics.class-ph]}
  \BibitemShut {NoStop}%
\bibitem [{\citenamefont {Hawking}\ and\ \citenamefont
  {Penrose}(1970)}]{Hawking:1970zqf}%
  \BibitemOpen
  \bibfield  {author} {\bibinfo {author} {\bibfnamefont {S.~W.}\ \bibnamefont
  {Hawking}}\ and\ \bibinfo {author} {\bibfnamefont {R.}~\bibnamefont
  {Penrose}},\ }\bibfield  {title} {\enquote {\bibinfo {title} {{The
  Singularities of gravitational collapse and cosmology}},}\ }\href {\doibase
  10.1098/rspa.1970.0021} {\bibfield  {journal} {\bibinfo  {journal} {Proc.
  Roy. Soc. Lond. A}\ }\textbf {\bibinfo {volume} {314}},\ \bibinfo {pages}
  {529--548} (\bibinfo {year} {1970})}\BibitemShut {NoStop}%
\bibitem [{\citenamefont {Ashtekar}\ and\ \citenamefont
  {Singh}(2011)}]{Ashtekar:2011ni}%
  \BibitemOpen
  \bibfield  {author} {\bibinfo {author} {\bibfnamefont {Abhay}\ \bibnamefont
  {Ashtekar}}\ and\ \bibinfo {author} {\bibfnamefont {Parampreet}\ \bibnamefont
  {Singh}},\ }\bibfield  {title} {\enquote {\bibinfo {title} {{Loop Quantum
  Cosmology: A Status Report}},}\ }\href {\doibase
  10.1088/0264-9381/28/21/213001} {\bibfield  {journal} {\bibinfo  {journal}
  {Class. Quant. Grav.}\ }\textbf {\bibinfo {volume} {28}},\ \bibinfo {pages}
  {213001} (\bibinfo {year} {2011})},\ \Eprint {http://arxiv.org/abs/1108.0893}
  {arXiv:1108.0893 [gr-qc]} \BibitemShut {NoStop}%
\bibitem [{\citenamefont {Oriti}(2006)}]{Oriti:2006se}%
  \BibitemOpen
  \bibfield  {author} {\bibinfo {author} {\bibfnamefont {Daniele}\ \bibnamefont
  {Oriti}},\ }\bibfield  {title} {\enquote {\bibinfo {title} {{The Group field
  theory approach to quantum gravity}},}\ }in\ \href@noop {} {\emph {\bibinfo
  {booktitle} {Approaches to Quantum Gravity - toward a new understanding of
  space, time, and matter}}},\ \bibinfo {editor} {edited by\ \bibinfo {editor}
  {\bibfnamefont {D.}~\bibnamefont {Oriti}}}\ (\bibinfo  {publisher} {Cambridge
  University Press},\ \bibinfo {year} {2006})\ \Eprint
  {http://arxiv.org/abs/gr-qc/0607032} {arXiv:gr-qc/0607032 [gr-qc]}
  \BibitemShut {NoStop}%
\bibitem [{\citenamefont {Oriti}\ \emph {et~al.}(2016)\citenamefont {Oriti},
  \citenamefont {Sindoni},\ and\ \citenamefont {Wilson-Ewing}}]{Oriti:2016qtz}%
  \BibitemOpen
  \bibfield  {author} {\bibinfo {author} {\bibfnamefont {Daniele}\ \bibnamefont
  {Oriti}}, \bibinfo {author} {\bibfnamefont {Lorenzo}\ \bibnamefont
  {Sindoni}}, \ and\ \bibinfo {author} {\bibfnamefont {Edward}\ \bibnamefont
  {Wilson-Ewing}},\ }\bibfield  {title} {\enquote {\bibinfo {title} {{Emergent
  Friedmann dynamics with a quantum bounce from quantum gravity
  condensates}},}\ }\href {\doibase 10.1088/0264-9381/33/22/224001} {\bibfield
  {journal} {\bibinfo  {journal} {Class. Quant. Grav.}\ }\textbf {\bibinfo
  {volume} {33}},\ \bibinfo {pages} {224001} (\bibinfo {year} {2016})},\
  \Eprint {http://arxiv.org/abs/1602.05881} {arXiv:1602.05881 [gr-qc]}
  \BibitemShut {NoStop}%
\bibitem [{\citenamefont {Oriti}(2024)}]{Oriti:2024elx}%
  \BibitemOpen
  \bibfield  {author} {\bibinfo {author} {\bibfnamefont {Daniele}\ \bibnamefont
  {Oriti}},\ }\bibfield  {title} {\enquote {\bibinfo {title} {{Hydrodynamics
  on~(Mini)superspace or~a~Non-linear Extension of~Quantum Cosmology: An
  Effective Timeless Framework for Cosmology from Quantum Gravity}},}\ }\href
  {\doibase 10.1007/978-3-031-61860-4_11} {\bibfield  {journal} {\bibinfo
  {journal} {Fundam. Theor. Phys.}\ }\textbf {\bibinfo {volume} {216}},\
  \bibinfo {pages} {221--252} (\bibinfo {year} {2024})}\BibitemShut {NoStop}%
\bibitem [{\citenamefont {Pithis}\ and\ \citenamefont
  {Sakellariadou}(2019)}]{Pithis:2019tvp}%
  \BibitemOpen
  \bibfield  {author} {\bibinfo {author} {\bibfnamefont {A.~G.~A.}\
  \bibnamefont {Pithis}}\ and\ \bibinfo {author} {\bibfnamefont
  {M.}~\bibnamefont {Sakellariadou}},\ }\bibfield  {title} {\enquote {\bibinfo
  {title} {{Group field theory condensate cosmology: An appetizer}},}\ }\href
  {\doibase 10.3390/universe5060147} {\bibfield  {journal} {\bibinfo  {journal}
  {Universe}\ }\textbf {\bibinfo {volume} {5}},\ \bibinfo {pages} {147}
  (\bibinfo {year} {2019})},\ \Eprint {http://arxiv.org/abs/1904.00598}
  {arXiv:1904.00598 [gr-qc]} \BibitemShut {NoStop}%
\bibitem [{\citenamefont {Gabbanelli}\ and\ \citenamefont
  {De~Bianchi}(2021)}]{Gabbanelli:2020lme}%
  \BibitemOpen
  \bibfield  {author} {\bibinfo {author} {\bibfnamefont {Luciano}\ \bibnamefont
  {Gabbanelli}}\ and\ \bibinfo {author} {\bibfnamefont {Silvia}\ \bibnamefont
  {De~Bianchi}},\ }\bibfield  {title} {\enquote {\bibinfo {title}
  {{Cosmological implications of the hydrodynamical phase of group field
  theory}},}\ }\href {\doibase 10.1007/s10714-021-02833-z} {\bibfield
  {journal} {\bibinfo  {journal} {Gen. Rel. Grav.}\ }\textbf {\bibinfo {volume}
  {53}},\ \bibinfo {pages} {66} (\bibinfo {year} {2021})},\ \Eprint
  {http://arxiv.org/abs/2008.07837} {arXiv:2008.07837 [gr-qc]} \BibitemShut
  {NoStop}%
\bibitem [{\citenamefont {Barca}\ \emph {et~al.}(2021)\citenamefont {Barca},
  \citenamefont {Giovannetti},\ and\ \citenamefont {Montani}}]{Barca:2021qdn}%
  \BibitemOpen
  \bibfield  {author} {\bibinfo {author} {\bibfnamefont {Gabriele}\
  \bibnamefont {Barca}}, \bibinfo {author} {\bibfnamefont {Eleonora}\
  \bibnamefont {Giovannetti}}, \ and\ \bibinfo {author} {\bibfnamefont
  {Giovanni}\ \bibnamefont {Montani}},\ }\bibfield  {title} {\enquote {\bibinfo
  {title} {{An Overview on the Nature of the Bounce in LQC and PQM}},}\ }\href
  {\doibase 10.3390/universe7090327} {\bibfield  {journal} {\bibinfo  {journal}
  {Universe}\ }\textbf {\bibinfo {volume} {7}},\ \bibinfo {pages} {327}
  (\bibinfo {year} {2021})},\ \Eprint {http://arxiv.org/abs/2109.08645}
  {arXiv:2109.08645 [gr-qc]} \BibitemShut {NoStop}%
\bibitem [{\citenamefont {Gasperini}\ and\ \citenamefont
  {Veneziano}(2003)}]{Gasperini:2002bn}%
  \BibitemOpen
  \bibfield  {author} {\bibinfo {author} {\bibfnamefont {M.}~\bibnamefont
  {Gasperini}}\ and\ \bibinfo {author} {\bibfnamefont {G.}~\bibnamefont
  {Veneziano}},\ }\bibfield  {title} {\enquote {\bibinfo {title} {{The Pre -
  big bang scenario in string cosmology}},}\ }\href {\doibase
  10.1016/S0370-1573(02)00389-7} {\bibfield  {journal} {\bibinfo  {journal}
  {Phys. Rept.}\ }\textbf {\bibinfo {volume} {373}},\ \bibinfo {pages} {1--212}
  (\bibinfo {year} {2003})},\ \Eprint {http://arxiv.org/abs/hep-th/0207130}
  {arXiv:hep-th/0207130} \BibitemShut {NoStop}%
\bibitem [{\citenamefont {Gasperini}\ and\ \citenamefont
  {Veneziano}(2023)}]{Gasperini:2023tus}%
  \BibitemOpen
  \bibfield  {author} {\bibinfo {author} {\bibfnamefont {M.}~\bibnamefont
  {Gasperini}}\ and\ \bibinfo {author} {\bibfnamefont {G.}~\bibnamefont
  {Veneziano}},\ }\bibfield  {title} {\enquote {\bibinfo {title} {{Non-singular
  pre-big bang scenarios from all-order \ensuremath{\alpha}' corrections}},}\
  }\href {\doibase 10.1007/JHEP07(2023)144} {\bibfield  {journal} {\bibinfo
  {journal} {JHEP}\ }\textbf {\bibinfo {volume} {07}},\ \bibinfo {pages} {144}
  (\bibinfo {year} {2023})},\ \Eprint {http://arxiv.org/abs/2305.00222}
  {arXiv:2305.00222 [hep-th]} \BibitemShut {NoStop}%
\bibitem [{\citenamefont {Veneziano}(1999)}]{Veneziano:1999ts}%
  \BibitemOpen
  \bibfield  {author} {\bibinfo {author} {\bibfnamefont {G.}~\bibnamefont
  {Veneziano}},\ }\bibfield  {title} {\enquote {\bibinfo {title} {{Pre -
  bangian origin of our entropy and time arrow}},}\ }\href {\doibase
  10.1016/S0370-2693(99)00267-1} {\bibfield  {journal} {\bibinfo  {journal}
  {Phys. Lett. B}\ }\textbf {\bibinfo {volume} {454}},\ \bibinfo {pages}
  {22--26} (\bibinfo {year} {1999})},\ \Eprint
  {http://arxiv.org/abs/hep-th/9902126} {arXiv:hep-th/9902126} \BibitemShut
  {NoStop}%
\bibitem [{\citenamefont {Conzinu}\ \emph {et~al.}(2023)\citenamefont
  {Conzinu}, \citenamefont {Fanizza}, \citenamefont {Gasperini}, \citenamefont
  {Pavone}, \citenamefont {Tedesco},\ and\ \citenamefont
  {Veneziano}}]{Conzinu:2023fth}%
  \BibitemOpen
  \bibfield  {author} {\bibinfo {author} {\bibfnamefont {P.}~\bibnamefont
  {Conzinu}}, \bibinfo {author} {\bibfnamefont {G.}~\bibnamefont {Fanizza}},
  \bibinfo {author} {\bibfnamefont {M.}~\bibnamefont {Gasperini}}, \bibinfo
  {author} {\bibfnamefont {E.}~\bibnamefont {Pavone}}, \bibinfo {author}
  {\bibfnamefont {L.}~\bibnamefont {Tedesco}}, \ and\ \bibinfo {author}
  {\bibfnamefont {G.}~\bibnamefont {Veneziano}},\ }\bibfield  {title} {\enquote
  {\bibinfo {title} {{From the string vacuum to FLRW or de Sitter via
  \ensuremath{\alpha}' corrections}},}\ }\href {\doibase
  10.1088/1475-7516/2023/12/019} {\bibfield  {journal} {\bibinfo  {journal}
  {JCAP}\ }\textbf {\bibinfo {volume} {12}},\ \bibinfo {pages} {019} (\bibinfo
  {year} {2023})},\ \Eprint {http://arxiv.org/abs/2308.16076} {arXiv:2308.16076
  [hep-th]} \BibitemShut {NoStop}%
\bibitem [{\citenamefont {Markov}(1982)}]{markov1982limiting}%
  \BibitemOpen
  \bibfield  {author} {\bibinfo {author} {\bibfnamefont {M.~A.}\ \bibnamefont
  {Markov}},\ }\bibfield  {title} {\enquote {\bibinfo {title} {Limiting density
  of matter as a universal law of nature},}\ }\href@noop {} {\bibfield
  {journal} {\bibinfo  {journal} {JETP Lett}\ }\textbf {\bibinfo {volume}
  {36}},\ \bibinfo {pages} {265} (\bibinfo {year} {1982})}\BibitemShut
  {NoStop}%
\bibitem [{\citenamefont {Mukhanov}\ and\ \citenamefont
  {Brandenberger}(1992)}]{Mukhanov:1991zn}%
  \BibitemOpen
  \bibfield  {author} {\bibinfo {author} {\bibfnamefont {Viatcheslav~F.}\
  \bibnamefont {Mukhanov}}\ and\ \bibinfo {author} {\bibfnamefont {Robert~H.}\
  \bibnamefont {Brandenberger}},\ }\bibfield  {title} {\enquote {\bibinfo
  {title} {{A Nonsingular universe}},}\ }\href {\doibase
  10.1103/PhysRevLett.68.1969} {\bibfield  {journal} {\bibinfo  {journal}
  {Phys. Rev. Lett.}\ }\textbf {\bibinfo {volume} {68}},\ \bibinfo {pages}
  {1969--1972} (\bibinfo {year} {1992})}\BibitemShut {NoStop}%
\bibitem [{\citenamefont {Easson}\ \emph {et~al.}(2011)\citenamefont {Easson},
  \citenamefont {Sawicki},\ and\ \citenamefont {Vikman}}]{Easson:2011zy}%
  \BibitemOpen
  \bibfield  {author} {\bibinfo {author} {\bibfnamefont {Damien~A.}\
  \bibnamefont {Easson}}, \bibinfo {author} {\bibfnamefont {Ignacy}\
  \bibnamefont {Sawicki}}, \ and\ \bibinfo {author} {\bibfnamefont {Alexander}\
  \bibnamefont {Vikman}},\ }\bibfield  {title} {\enquote {\bibinfo {title}
  {{G-Bounce}},}\ }\href {\doibase 10.1088/1475-7516/2011/11/021} {\bibfield
  {journal} {\bibinfo  {journal} {JCAP}\ }\textbf {\bibinfo {volume} {11}},\
  \bibinfo {pages} {021} (\bibinfo {year} {2011})},\ \Eprint
  {http://arxiv.org/abs/1109.1047} {arXiv:1109.1047 [hep-th]} \BibitemShut
  {NoStop}%
\bibitem [{\citenamefont {Cai}\ \emph {et~al.}(2012{\natexlab{a}})\citenamefont
  {Cai}, \citenamefont {Easson},\ and\ \citenamefont
  {Brandenberger}}]{Cai:2012va}%
  \BibitemOpen
  \bibfield  {author} {\bibinfo {author} {\bibfnamefont {Yi-Fu}\ \bibnamefont
  {Cai}}, \bibinfo {author} {\bibfnamefont {Damien~A.}\ \bibnamefont {Easson}},
  \ and\ \bibinfo {author} {\bibfnamefont {Robert}\ \bibnamefont
  {Brandenberger}},\ }\bibfield  {title} {\enquote {\bibinfo {title} {{Towards
  a Nonsingular Bouncing Cosmology}},}\ }\href {\doibase
  10.1088/1475-7516/2012/08/020} {\bibfield  {journal} {\bibinfo  {journal}
  {JCAP}\ }\textbf {\bibinfo {volume} {08}},\ \bibinfo {pages} {020} (\bibinfo
  {year} {2012}{\natexlab{a}})},\ \Eprint {http://arxiv.org/abs/1206.2382}
  {arXiv:1206.2382 [hep-th]} \BibitemShut {NoStop}%
\bibitem [{\citenamefont {Cai}\ \emph {et~al.}(2012{\natexlab{b}})\citenamefont
  {Cai}, \citenamefont {Gao},\ and\ \citenamefont {Saridakis}}]{Cai:2012ag}%
  \BibitemOpen
  \bibfield  {author} {\bibinfo {author} {\bibfnamefont {Yi-Fu}\ \bibnamefont
  {Cai}}, \bibinfo {author} {\bibfnamefont {Caixia}\ \bibnamefont {Gao}}, \
  and\ \bibinfo {author} {\bibfnamefont {Emmanuel~N.}\ \bibnamefont
  {Saridakis}},\ }\bibfield  {title} {\enquote {\bibinfo {title} {{Bounce and
  cyclic cosmology in extended nonlinear massive gravity}},}\ }\href {\doibase
  10.1088/1475-7516/2012/10/048} {\bibfield  {journal} {\bibinfo  {journal}
  {JCAP}\ }\textbf {\bibinfo {volume} {10}},\ \bibinfo {pages} {048} (\bibinfo
  {year} {2012}{\natexlab{b}})},\ \Eprint {http://arxiv.org/abs/1207.3786}
  {arXiv:1207.3786 [astro-ph.CO]} \BibitemShut {NoStop}%
\bibitem [{\citenamefont {Ijjas}\ and\ \citenamefont
  {Steinhardt}(2016)}]{Ijjas:2016tpn}%
  \BibitemOpen
  \bibfield  {author} {\bibinfo {author} {\bibfnamefont {Anna}\ \bibnamefont
  {Ijjas}}\ and\ \bibinfo {author} {\bibfnamefont {Paul~J.}\ \bibnamefont
  {Steinhardt}},\ }\bibfield  {title} {\enquote {\bibinfo {title} {{Classically
  stable nonsingular cosmological bounces}},}\ }\href {\doibase
  10.1103/PhysRevLett.117.121304} {\bibfield  {journal} {\bibinfo  {journal}
  {Phys. Rev. Lett.}\ }\textbf {\bibinfo {volume} {117}},\ \bibinfo {pages}
  {121304} (\bibinfo {year} {2016})},\ \Eprint
  {http://arxiv.org/abs/1606.08880} {arXiv:1606.08880 [gr-qc]} \BibitemShut
  {NoStop}%
\bibitem [{\citenamefont {Chamseddine}\ and\ \citenamefont
  {Mukhanov}(2017)}]{Chamseddine:2016uef}%
  \BibitemOpen
  \bibfield  {author} {\bibinfo {author} {\bibfnamefont {Ali~H.}\ \bibnamefont
  {Chamseddine}}\ and\ \bibinfo {author} {\bibfnamefont {Viatcheslav}\
  \bibnamefont {Mukhanov}},\ }\bibfield  {title} {\enquote {\bibinfo {title}
  {{Resolving Cosmological Singularities}},}\ }\href {\doibase
  10.1088/1475-7516/2017/03/009} {\bibfield  {journal} {\bibinfo  {journal}
  {JCAP}\ }\textbf {\bibinfo {volume} {03}},\ \bibinfo {pages} {009} (\bibinfo
  {year} {2017})},\ \Eprint {http://arxiv.org/abs/1612.05860} {arXiv:1612.05860
  [gr-qc]} \BibitemShut {NoStop}%
\bibitem [{\citenamefont {de~Cesare}(2019{\natexlab{a}})}]{deCesare:2018cts}%
  \BibitemOpen
  \bibfield  {author} {\bibinfo {author} {\bibfnamefont {Marco}\ \bibnamefont
  {de~Cesare}},\ }\bibfield  {title} {\enquote {\bibinfo {title} {{Limiting
  curvature mimetic gravity for group field theory condensates}},}\ }\href
  {\doibase 10.1103/PhysRevD.99.063505} {\bibfield  {journal} {\bibinfo
  {journal} {Phys. Rev. D}\ }\textbf {\bibinfo {volume} {99}},\ \bibinfo
  {pages} {063505} (\bibinfo {year} {2019}{\natexlab{a}})},\ \Eprint
  {http://arxiv.org/abs/1812.06171} {arXiv:1812.06171 [gr-qc]} \BibitemShut
  {NoStop}%
\bibitem [{\citenamefont {de~Cesare}\ and\ \citenamefont
  {Wilson-Ewing}(2019)}]{deCesare:2019suk}%
  \BibitemOpen
  \bibfield  {author} {\bibinfo {author} {\bibfnamefont {Marco}\ \bibnamefont
  {de~Cesare}}\ and\ \bibinfo {author} {\bibfnamefont {Edward}\ \bibnamefont
  {Wilson-Ewing}},\ }\bibfield  {title} {\enquote {\bibinfo {title} {{A
  generalized Kasner transition for bouncing Bianchi I models in modified
  gravity theories}},}\ }\href {\doibase 10.1088/1475-7516/2019/12/039}
  {\bibfield  {journal} {\bibinfo  {journal} {JCAP}\ }\textbf {\bibinfo
  {volume} {12}},\ \bibinfo {pages} {039} (\bibinfo {year} {2019})},\ \Eprint
  {http://arxiv.org/abs/1910.03616} {arXiv:1910.03616 [gr-qc]} \BibitemShut
  {NoStop}%
\bibitem [{\citenamefont {Ilyas}\ \emph {et~al.}(2020)\citenamefont {Ilyas},
  \citenamefont {Zhu}, \citenamefont {Zheng}, \citenamefont {Cai},\ and\
  \citenamefont {Saridakis}}]{Ilyas:2020qja}%
  \BibitemOpen
  \bibfield  {author} {\bibinfo {author} {\bibfnamefont {Amara}\ \bibnamefont
  {Ilyas}}, \bibinfo {author} {\bibfnamefont {Mian}\ \bibnamefont {Zhu}},
  \bibinfo {author} {\bibfnamefont {Yunlong}\ \bibnamefont {Zheng}}, \bibinfo
  {author} {\bibfnamefont {Yi-Fu}\ \bibnamefont {Cai}}, \ and\ \bibinfo
  {author} {\bibfnamefont {Emmanuel~N.}\ \bibnamefont {Saridakis}},\ }\bibfield
   {title} {\enquote {\bibinfo {title} {{DHOST Bounce}},}\ }\href {\doibase
  10.1088/1475-7516/2020/09/002} {\bibfield  {journal} {\bibinfo  {journal}
  {JCAP}\ }\textbf {\bibinfo {volume} {09}},\ \bibinfo {pages} {002} (\bibinfo
  {year} {2020})},\ \Eprint {http://arxiv.org/abs/2002.08269} {arXiv:2002.08269
  [gr-qc]} \BibitemShut {NoStop}%
\bibitem [{\citenamefont {de~Cesare}(2019{\natexlab{b}})}]{deCesare:2019pqj}%
  \BibitemOpen
  \bibfield  {author} {\bibinfo {author} {\bibfnamefont {Marco}\ \bibnamefont
  {de~Cesare}},\ }\bibfield  {title} {\enquote {\bibinfo {title}
  {{Reconstruction of Mimetic Gravity in a Non-SingularBouncing Universe from
  Quantum Gravity}},}\ }\href {\doibase 10.3390/universe5050107} {\bibfield
  {journal} {\bibinfo  {journal} {Universe}\ }\textbf {\bibinfo {volume} {5}},\
  \bibinfo {pages} {107} (\bibinfo {year} {2019}{\natexlab{b}})},\ \Eprint
  {http://arxiv.org/abs/1904.02622} {arXiv:1904.02622 [gr-qc]} \BibitemShut
  {NoStop}%
\bibitem [{\citenamefont {Brandenberger}\ and\ \citenamefont
  {Peter}(2017)}]{Brandenberger:2016vhg}%
  \BibitemOpen
  \bibfield  {author} {\bibinfo {author} {\bibfnamefont {Robert}\ \bibnamefont
  {Brandenberger}}\ and\ \bibinfo {author} {\bibfnamefont {Patrick}\
  \bibnamefont {Peter}},\ }\bibfield  {title} {\enquote {\bibinfo {title}
  {{Bouncing Cosmologies: Progress and Problems}},}\ }\href {\doibase
  10.1007/s10701-016-0057-0} {\bibfield  {journal} {\bibinfo  {journal} {Found.
  Phys.}\ }\textbf {\bibinfo {volume} {47}},\ \bibinfo {pages} {797--850}
  (\bibinfo {year} {2017})},\ \Eprint {http://arxiv.org/abs/1603.05834}
  {arXiv:1603.05834 [hep-th]} \BibitemShut {NoStop}%
\bibitem [{\citenamefont {Agullo}\ \emph {et~al.}(2023)\citenamefont {Agullo},
  \citenamefont {Wang},\ and\ \citenamefont {Wilson-Ewing}}]{Agullo:2023rqq}%
  \BibitemOpen
  \bibfield  {author} {\bibinfo {author} {\bibfnamefont {Ivan}\ \bibnamefont
  {Agullo}}, \bibinfo {author} {\bibfnamefont {Anzhong}\ \bibnamefont {Wang}},
  \ and\ \bibinfo {author} {\bibfnamefont {Edward}\ \bibnamefont
  {Wilson-Ewing}},\ }\enquote {\bibinfo {title} {Loop quantum cosmology:
  Relation between theory and observations},}\ in\ \href {\doibase
  10.1007/978-981-19-3079-9_103-1} {\emph {\bibinfo {booktitle} {Handbook of
  Quantum Gravity}}}\ (\bibinfo  {publisher} {Springer Nature Singapore},\
  \bibinfo {year} {2023})\ pp.\ \bibinfo {pages} {1--46},\ \Eprint
  {http://arxiv.org/abs/2301.10215} {arXiv:2301.10215 [gr-qc]} \BibitemShut
  {NoStop}%
\bibitem [{\citenamefont {Singh}(2009)}]{Singh:2009mz}%
  \BibitemOpen
  \bibfield  {author} {\bibinfo {author} {\bibfnamefont {Parampreet}\
  \bibnamefont {Singh}},\ }\bibfield  {title} {\enquote {\bibinfo {title} {{Are
  loop quantum cosmos never singular?}}}\ }\href {\doibase
  10.1088/0264-9381/26/12/125005} {\bibfield  {journal} {\bibinfo  {journal}
  {Class. Quant. Grav.}\ }\textbf {\bibinfo {volume} {26}},\ \bibinfo {pages}
  {125005} (\bibinfo {year} {2009})},\ \Eprint {http://arxiv.org/abs/0901.2750}
  {arXiv:0901.2750 [gr-qc]} \BibitemShut {NoStop}%
\bibitem [{\citenamefont {Cai}\ and\ \citenamefont
  {Wilson-Ewing}(2015)}]{Cai:2014jla}%
  \BibitemOpen
  \bibfield  {author} {\bibinfo {author} {\bibfnamefont {Yi-Fu}\ \bibnamefont
  {Cai}}\ and\ \bibinfo {author} {\bibfnamefont {Edward}\ \bibnamefont
  {Wilson-Ewing}},\ }\bibfield  {title} {\enquote {\bibinfo {title} {{A
  $\Lambda$CDM bounce scenario}},}\ }\href {\doibase
  10.1088/1475-7516/2015/03/006} {\bibfield  {journal} {\bibinfo  {journal}
  {JCAP}\ }\textbf {\bibinfo {volume} {03}},\ \bibinfo {pages} {006} (\bibinfo
  {year} {2015})},\ \Eprint {http://arxiv.org/abs/1412.2914} {arXiv:1412.2914
  [gr-qc]} \BibitemShut {NoStop}%
\bibitem [{\citenamefont {Cai}\ \emph {et~al.}(2016)\citenamefont {Cai},
  \citenamefont {Marciano}, \citenamefont {Wang},\ and\ \citenamefont
  {Wilson-Ewing}}]{Cai:2016hea}%
  \BibitemOpen
  \bibfield  {author} {\bibinfo {author} {\bibfnamefont {Yi-Fu}\ \bibnamefont
  {Cai}}, \bibinfo {author} {\bibfnamefont {Antonino}\ \bibnamefont
  {Marciano}}, \bibinfo {author} {\bibfnamefont {Dong-Gang}\ \bibnamefont
  {Wang}}, \ and\ \bibinfo {author} {\bibfnamefont {Edward}\ \bibnamefont
  {Wilson-Ewing}},\ }\bibfield  {title} {\enquote {\bibinfo {title} {{Bouncing
  cosmologies with dark matter and dark energy}},}\ }\href {\doibase
  10.3390/universe3010001} {\bibfield  {journal} {\bibinfo  {journal}
  {Universe}\ }\textbf {\bibinfo {volume} {3}},\ \bibinfo {pages} {1} (\bibinfo
  {year} {2016})},\ \Eprint {http://arxiv.org/abs/1610.00938} {arXiv:1610.00938
  [astro-ph.CO]} \BibitemShut {NoStop}%
\bibitem [{\citenamefont {Wilson-Ewing}(2013)}]{Wilson-Ewing:2012lmx}%
  \BibitemOpen
  \bibfield  {author} {\bibinfo {author} {\bibfnamefont {Edward}\ \bibnamefont
  {Wilson-Ewing}},\ }\bibfield  {title} {\enquote {\bibinfo {title} {{The
  Matter Bounce Scenario in Loop Quantum Cosmology}},}\ }\href {\doibase
  10.1088/1475-7516/2013/03/026} {\bibfield  {journal} {\bibinfo  {journal}
  {JCAP}\ }\textbf {\bibinfo {volume} {03}},\ \bibinfo {pages} {026} (\bibinfo
  {year} {2013})},\ \Eprint {http://arxiv.org/abs/1211.6269} {arXiv:1211.6269
  [gr-qc]} \BibitemShut {NoStop}%
\bibitem [{\citenamefont {Chinaglia}\ \emph {et~al.}(2017)\citenamefont
  {Chinaglia}, \citenamefont {Coll\'eaux},\ and\ \citenamefont
  {Zerbini}}]{Chinaglia:2017wim}%
  \BibitemOpen
  \bibfield  {author} {\bibinfo {author} {\bibfnamefont {Stefano}\ \bibnamefont
  {Chinaglia}}, \bibinfo {author} {\bibfnamefont {Aimeric}\ \bibnamefont
  {Coll\'eaux}}, \ and\ \bibinfo {author} {\bibfnamefont {Sergio}\ \bibnamefont
  {Zerbini}},\ }\bibfield  {title} {\enquote {\bibinfo {title} {{A
  non-polynomial gravity formulation for Loop Quantum Cosmology bounce}},}\
  }\href {\doibase 10.3390/galaxies5030051} {\bibfield  {journal} {\bibinfo
  {journal} {Galaxies}\ }\textbf {\bibinfo {volume} {5}},\ \bibinfo {pages}
  {51} (\bibinfo {year} {2017})},\ \Eprint {http://arxiv.org/abs/1708.08667}
  {arXiv:1708.08667 [gr-qc]} \BibitemShut {NoStop}%
\bibitem [{\citenamefont {Faraoni}(2015)}]{Faraoni_2015}%
  \BibitemOpen
  \bibfield  {author} {\bibinfo {author} {\bibfnamefont {Valerio}\ \bibnamefont
  {Faraoni}},\ }\href {\doibase 10.1007/978-3-319-19240-6} {\emph {\bibinfo
  {title} {Cosmological and Black Hole Apparent Horizons}}}\ (\bibinfo
  {publisher} {Springer International Publishing},\ \bibinfo {year}
  {2015})\BibitemShut {NoStop}%
\bibitem [{\citenamefont {Quintin}\ and\ \citenamefont
  {Brandenberger}(2016)}]{Quintin:2016qro}%
  \BibitemOpen
  \bibfield  {author} {\bibinfo {author} {\bibfnamefont {Jerome}\ \bibnamefont
  {Quintin}}\ and\ \bibinfo {author} {\bibfnamefont {Robert~H.}\ \bibnamefont
  {Brandenberger}},\ }\bibfield  {title} {\enquote {\bibinfo {title} {{Black
  hole formation in a contracting universe}},}\ }\href {\doibase
  10.1088/1475-7516/2016/11/029} {\bibfield  {journal} {\bibinfo  {journal}
  {JCAP}\ }\textbf {\bibinfo {volume} {11}},\ \bibinfo {pages} {029} (\bibinfo
  {year} {2016})},\ \Eprint {http://arxiv.org/abs/1609.02556} {arXiv:1609.02556
  [astro-ph.CO]} \BibitemShut {NoStop}%
\bibitem [{\citenamefont {Barrow}(1987)}]{Barrow:1987ia}%
  \BibitemOpen
  \bibfield  {author} {\bibinfo {author} {\bibfnamefont {John~D.}\ \bibnamefont
  {Barrow}},\ }\bibfield  {title} {\enquote {\bibinfo {title} {{Cosmic No Hair
  Theorems and Inflation}},}\ }\href {\doibase 10.1016/0370-2693(87)90063-3}
  {\bibfield  {journal} {\bibinfo  {journal} {Phys. Lett. B}\ }\textbf
  {\bibinfo {volume} {187}},\ \bibinfo {pages} {12--16} (\bibinfo {year}
  {1987})}\BibitemShut {NoStop}%
\bibitem [{\citenamefont {Erickson}\ \emph {et~al.}(2004)\citenamefont
  {Erickson}, \citenamefont {Wesley}, \citenamefont {Steinhardt},\ and\
  \citenamefont {Turok}}]{Erickson:2003zm}%
  \BibitemOpen
  \bibfield  {author} {\bibinfo {author} {\bibfnamefont {Joel~K.}\ \bibnamefont
  {Erickson}}, \bibinfo {author} {\bibfnamefont {Daniel~H.}\ \bibnamefont
  {Wesley}}, \bibinfo {author} {\bibfnamefont {Paul~J.}\ \bibnamefont
  {Steinhardt}}, \ and\ \bibinfo {author} {\bibfnamefont {Neil}\ \bibnamefont
  {Turok}},\ }\bibfield  {title} {\enquote {\bibinfo {title} {{Kasner and
  mixmaster behavior in universes with equation of state w \ensuremath{>}=
  1}},}\ }\href {\doibase 10.1103/PhysRevD.69.063514} {\bibfield  {journal}
  {\bibinfo  {journal} {Phys. Rev. D}\ }\textbf {\bibinfo {volume} {69}},\
  \bibinfo {pages} {063514} (\bibinfo {year} {2004})},\ \Eprint
  {http://arxiv.org/abs/hep-th/0312009} {arXiv:hep-th/0312009} \BibitemShut
  {NoStop}%
\bibitem [{\citenamefont {V.}(2011)}]{KUCHA__2011}%
  \BibitemOpen
  \bibfield  {author} {\bibinfo {author} {\bibfnamefont {Kucha\v{r}~Karel}\
  \bibnamefont {V.}},\ }\bibfield  {title} {\enquote {\bibinfo {title} {Time
  and interpretations of quantum gravity},}\ }\href {\doibase
  10.1142/s0218271811019347} {\bibfield  {journal} {\bibinfo  {journal}
  {International Journal of Modern Physics D}\ }\textbf {\bibinfo {volume}
  {20}},\ \bibinfo {pages} {3--86} (\bibinfo {year} {2011})}\BibitemShut
  {NoStop}%
\bibitem [{\citenamefont {Forgione}(2024)}]{Forgione:2024upw}%
  \BibitemOpen
  \bibfield  {author} {\bibinfo {author} {\bibfnamefont {Marco}\ \bibnamefont
  {Forgione}},\ }\bibfield  {title} {\enquote {\bibinfo {title} {{Causation as
  Constraints in Causal Set Theory}},}\ }\href {\doibase
  10.1007/978-3-031-61860-4_6} {\bibfield  {journal} {\bibinfo  {journal}
  {Fundam. Theor. Phys.}\ }\textbf {\bibinfo {volume} {216}},\ \bibinfo {pages}
  {107--125} (\bibinfo {year} {2024})}\BibitemShut {NoStop}%
\bibitem [{\citenamefont {De~Bianchi}\ and\ \citenamefont
  {Gabbanelli}(2023)}]{DeBianchi:2023yys}%
  \BibitemOpen
  \bibfield  {author} {\bibinfo {author} {\bibfnamefont {Silvia}\ \bibnamefont
  {De~Bianchi}}\ and\ \bibinfo {author} {\bibfnamefont {Luciano}\ \bibnamefont
  {Gabbanelli}},\ }\bibfield  {title} {\enquote {\bibinfo {title} {{Re-thinking
  geometrogenesis: Instantaneity in quantum gravity scenarios}},}\ }\href
  {\doibase 10.1088/1742-6596/2533/1/012001} {\bibfield  {journal} {\bibinfo
  {journal} {J. Phys. Conf. Ser.}\ }\textbf {\bibinfo {volume} {2533}},\
  \bibinfo {pages} {012001} (\bibinfo {year} {2023})}\BibitemShut {NoStop}%
\end{thebibliography}%

\end{document}